# A Multi-factor Multi-level and Interaction based (M2I) Authentication Framework for Internet of Things (IoT) Applications

**SALEM ALJANAH** [1], **NING ZHANG** [1], **AND SIOK WAH TAY** [1]
[1]Department of Computer Science, The University of Manchester, Manchester M13 9PL, U.K.

Corresponding author: Salem AlJanah (e-mail: salem.aljanah@manchester.ac.uk).

**ABSTRACT** Existing authentication solutions proposed for Internet of Things (IoT) provide a single Level of Assurance (LoA) regardless of the sensitivity levels of the resources or interactions between IoT devices being protected. For effective (with adequate level of protection) and efficient (with as low overhead costs as possible) protections, it may be desirable to tailor the protection level in response to the sensitivity level of the resources, as a stronger protection level typically imposes a higher level of overheads costs. In this paper, we investigate how to facilitate multi-LoA authentication for IoT by proposing a multi-factor multi-level and interaction based (M2I) authentication framework. The framework implements LoA linked and interaction based authentication. Two interaction modes are investigated, P2P (Peer-to-Peer) and O2M (One-to-Many) via the design of two corresponding protocols. Evaluation results show that adopting the O2M interaction mode in authentication can cut communication cost significantly; compared with that of the Kerberos protocol, the O2M protocol reduces the communication cost by $42\% \sim 45\%$. The protocols also introduce less computational cost. The P2P and O2M protocol, respectively, reduce the computational cost by $70\% \sim 72\%$ and $81\% \sim 82\%$ in comparison with that of Kerberos. Evaluation results also show that the two factor authentication option costs twice as much as that of the one-factor option.

## I. INTRODUCTION

The recent increase in the number of smart devices (i.e., devices that are capable of performing some communication and computational tasks autonomously [1]) has made a number of Internet of Things (IoT) applications, e.g., smart home, smart health, and industrial IoT, popular [2]. The use of these applications can help automate routine tasks, such as turning off the lights when it is daytime.

Although task automation may bring some convenience, several studies [3–5] have shown that it may also introduce a number of security challenges. One of the challenges is how to achieve effective and efficient authentication in an IoT environment where devices are heterogeneous [6], and some could have resource constraints such as limited processing power. By effective, we mean that the authentication service should be secure in authenticating heterogeneous and resource-constrained devices, and by efficient, we mean that the service should introduce as less overheads as possible.

To achieve effective and efficient authentication, attributes (e.g., asset value, location, and mode of interactions) that may influence the required level of protection (i.e., required LoA) may need to be considered so that more valuable assets and/or accessing them from a riskier location or a more security-sensitive interaction should be protected with an authentication method providing a stronger level of protection, and vice versa. A stronger level of protection is typically accompanied with a higher level of overhead costs, this multi-level approach to authentication may reduce unnecessary overhead costs while providing an adequate level of protection, optimising the trade-off between protection strengths and costs incurred in providing the protection. In evaluating this approach, we seek to answer the following research questions:

**FQ1:** How to facilitate multi-level (multiple levels of assurance, or multi-LoA) device-to-device authentication?

**FQ2:** How to minimise costs while facilitating the multi-level authentication?

**FQ3:** What is the effectiveness of the approach?

**FQ4:** What are the costs incurred in adopting the approach?

To scope the work without losing generality, we have carried out the research work using a smart home (SHome)

use-case as the underlying application context. An SHome typically hosts a variety of IoT devices and applications [7]. Hence, research outcome or any lessons learned should be applicable to other IoT based applications.

To investigate and evaluate the multi-LoA approach to device-to-device authentication in an IoT context, this paper examines how to quantify LoA and use it to govern how authentication should be carried out at run-time in an SHome environment, and proposes a multi-factor multi-level and interaction based (M2I) authentication framework. The framework consists of a required LoA (RLoA) method, three LoA derivation and aggregation methods, the LoA derivation (LoAD) method, the client derived LoA aggregation (CDLoA) method and the multi-client derived LoA aggregation (MCDLoA) method, and two authentication protocols, the Peer-to-Peer (P2P) and the One-to-Many (O2M) protocol. The RLoA method is used to determine the LoA needed to access a device. The LoAD method derives the LoA achieved by a user or a user device in an authentication instance. The CDLoA and MCDLoA method are used to aggregate the level of assurance values achieved by different authentication instances in a session of a single client or multiple clients, respectively. The P2P and O2M authentication protocol, referred to as the M2I protocols respectively support multi-factor and multi-LoA authentication of devices in device-to-device and device-to-multiDevice modes of interactions. The paper also presents both theoretical and experimental evaluations of the protocols with regard to their effectiveness and efficiency and compares the performance with that of the most related protocol.

The rest of the paper is organised as follows. Section II surveys related work. Section III discusses the high-level ideas used to design and evaluate the M2I framework. Section IV analyses the level of assurance required and how it may be derived in an SHome environment. Section V introduces design preliminaries. Section VI presents the M2I authentication framework. Section VII analyses the M2I protocols. Section VIII evaluates the protocols using experiments and discusses the experimental results. Section IX analyses the most related solution, the Kerberos version 5 protocol, and then compares the communication and computational cost of the M2I protocols with that of Kerberos. Finally, Section X concludes the paper.

## II. RELATED WORK

A number of architectures have been proposed to facilitate authentication in IoT applications. Some of these architectures are discussed below.

To enhance security, Amraoui et al. [8] proposed a machine learning based architecture to facilitate implicit and continuous authentication in the SHome environment. Tantidham and Aung [9] proposed a blockchain based architecture to secure communications between the SHome devices and external untrusted entities. Saadeh et al. [10] proposed asymmetric key-based multi-layer architecture to secure authentication in smart cities. Although these solutions may enhance security, they impose additional authentication overheads.

To reduce the authentication overhead, a number of hardware based authentication architectures have been proposed. Chatterjee et al. [11] proposed a Physical Unclonable Function (PUF) based architecture to secure authentication. Gope at al. [12] proposed a radio frequency identification (RFID) based architecture to facilitate authentication in distributed IoT environments. Although hardware based solutions may reduce the authentication overhead, they typically require clients to have additional hardware, such as a PUF circuit or an RFID tag [13].

A number of authentication protocols [12, 14–36] have also been proposed for IoT applications. In [13], we analysed these protocols and found that they provide a single LoA. In other words, they provide the same level of protection for different entities or interactions. Existing multi-LoA authentication solutions are often designed for user authentication; they are not suitable for device authentication, especially when the devices involved may have a number of constraints, e.g., limited processing capability, as in the case of IoT environments.

Even though some of the proposed solutions (i.e., authentication architectures, authentication methods and protocols) have advanced in securing IoT environments, there is still work to be done. For instance, how to optimise the trade-off between security and authentication overhead. One way to do this is by providing an authentication service that can adapt the level of protection offered by the service in adaptation to the level of assurance required to protect the action for which the authentication is performed.

## III. HIGH-LEVEL IDEAS AND THEIR IMPLEMENTATIONS

To reduce cost, we use the following ideas:

1) LoA linked authentication, i.e., to tailor authentication method in adaptation to the sensitivity level of the devices accessed.
2) interaction-based key sharing, i.e., authentication with all the devices in the group is done by using the same ticket (containing the same key).

The implementations of the ideas are as follows.

(i) **Ideas used for addressing FQ1 (How to facilitate multi-level device-to-device authentication?):**

To facilitate the multi-level authentication, the proposed framework should

- Support the use of multiple authentication factors. With the use of different factors or different groups of factors, a different level of authentication assurance can be achieved.
- Support LoA based authentication decision making.
  - Assign each resource-hosting device a required level of assurance value which represents the LoA

- needed to access the device or resources hosted or managed by the device.
- Assign each authentication factor a LoA value, and if two or more factors were used in a session, then the framework derives an aggregated LoA (Agg-DLoA) upon successful authentication.
- Grant access if the Agg-DloA value of the session is greater than or equal to the RLoA value of the target device. Otherwise, access is denied.

(ii) **Ideas for addressing FQ2 (How to minimise costs while facilitating the multi-level authentication?):**

The ideas and measures used to minimise costs incurred in authentication are as follows.
- Use LoA-based decision making to balance the trade-off between the level of protection and the level of cost.
- Allow interaction based authentication, where devices are authenticated according to their mode of interaction using different protocols, to reduce the number of tokens issued and verified, and the number of interactions in an authentication instance.
- Maximize the use of computationally efficient algorithms, e.g., symmetric ciphers.
- Use hash chain-based verification scheme to reduce the cost of future re-authentication.

## IV. LEVEL OF ASSURANCE

In an SHome environment, a resource (data or services) access is typically accomplished via the access to the device hosting or managing the resource. Depending on their roles, SHome devices (i.e., IoT devices that are hosted in an SHome environment) can be classified into two groups: a resource group and a user group. Depending on the sensitivity levels of the resources they host, each resource-hosting device (i.e., a device in the resource group) is assigned with a Required Levels of Assurance (RLoA). A RLoA value for a device is assumed to be determined prior to run-time and via risk assessment. Similarly, depending on the resource it accesses, each user device is required to have a Derived Level of Assurance (DLoA), and the DLoA value for a device is calculated at run-time. Multiple authentication factors may be involved in the derivation of a DLoA value. For the sake of clarity, we use DLoA to denote the assurance level derived by using a single authentication factor, and an aggregate DLoA (or Agg-DLoA) to denote the assurance level derived by using multiple authentication factors.

### A. REQUIRED LEVEL OF ASSURANCE (RLOA)

A RLoA value for a resource-hosting device represents the LoA needed to access the device or resources hosted or managed by the device. A class LoA (CLoA) represents the LoA value of a device. The CLoA value can be determined by a number of attributes, e.g., device capability (dc), asset value (av), and location (loc). Some attributes may be set during the registration phase, whereas others may be left to the SHome owner as their values may be a subjective matter. For instance, the dc value captures the device capability and hence can be set during registration. However, the av and loc values can be subjective and therefore are set by the owner of the SHome via policy specification. Table 1 and Table 2 describe exemplar settings of three levels of CLoA_dc and CLoA_av.

TABLE 1: An exemplar setting of $CLoA\_dc$

| CLoA_dc | Description |
|---|---|
| 1 | May not act as a client nor a proxy |
| 2 | May act as a client but not a proxy |
| 3 | May act as a client and a proxy |

TABLE 2: An exemplar setting of $CLoA\_av$

| CLoA_av | Description |
|---|---|
| 1 | Little or no value in the asset |
| 2 | The asset has some value |
| 3 | The asset has a high value |

**The RLoA Method**

If a target device has multiple LoA-effecting attributes and each has a CLoA value, then the RLoA for accessing the device should be equal to the highest CLoA value. In other words, given that a device has three CLoA effecting attributes and the CLoA values of these attributes are, respectively, CLoA_dc, CLoA_av, CLoA_loc, then the RLoA for accessing this device should be determined by the following equation.

$$RLoA = MAX(CLoA\_dc, CLoA\_av, CLoA\_loc) \quad (1)$$

For example, in the case of opening a safe, if the safe (with regard to device capability) has a CLoA_dc=2, but the value of the asset inside the safe (a lot of money inside) dictates that CLoA_av=3, then the RLoA should be the max of the two values, i.e., 3.

### B. DERIVED LEVEL OF ASSURANCE (DLOA)

A DLoA value of an authentication instance represents the LoA achieved by a user or a user device in that instance. A DLoA value is typically affected by a number of factors, such as the levels of assurance of the underlying authentication methods used, the trust levels of the respective authentication servers (reflected by their respective weightings), and the relationship between these factors. The DLoA of an authentication instance is calculated at run-time. If two or more authentication factors or methods are used, then an Aggregated DLoA (Agg-DLoA) should be calculated. Depending on the number of authentication instances and clients involved, the aggregation may be done using LoAD, CDLoA or MCDLoA method. Once aggregated, if the Agg-DLoA value is a fraction, the integer that is smaller than the actual value will be chosen.

**(i) The Level of Assurance Derivation (LoAD) Method**
If multiple authentication methods are used to verify the client identity in an authentication instance, then the weighted sum approach applies. In other words, given that a client used n authentication methods to verify its identity and the LoA values of these methods are $LoA_{AuthMethod_1}, \ldots LoA_{AuthMethod_n}$, then the Agg-DLoA of this instance should be determined by the following equation.

$$Agg - DLoA_{instance} = \sum_{i=1}^{n} W_{AuthMethod_i} \times LoA_{AuthMethod_i} \quad (2)$$

An example is if a username and password authentication method is used with an out-of-band authentication method (e.g., SMS) to verify the client identity, then the Agg-DLoA should be the weighted sum of the LoA values provided by these methods (i.e., $W_{UsernameAndPassword} \times LoA_{UsernameAndPassword} + W_{SMS} \times LoA_{SMS}$).

If a session has more than one instance, then the session Agg-DLoA ($Agg - DLoA_{session}$) should be calculated. The $Agg - DLoA_{session}$ value represents the overall assurance level of different authentication instances in the session. Depending on the number of clients, the $Agg - DLoA_{session}$ may be derived using the maximum or the weakest-link approach.

**(ii) The Client Derived Level of Assurance Aggregation (CD-LoA) Method**
If the session has one client (i.e., all authentication requests are made by the client itself and not through other devices), then the maximum approach applies. In other words, the $Agg - DLoA_{session}$ value should be equal to the highest Agg-DLoA value in the session.

$$Agg - DLoA_{session} = MAX(Agg - DLoA_1, Agg - DLoA_2,$$
$$\ldots\ldots\ldots\ldots\ldots Agg - DLoA_n) \quad (3)$$

For example, if a session has two different authentication instances initiated by the same client, and their Agg-DLoA values are 1 and 3, then the $Agg - DLoA_{session}$ should be the max of the two values, i.e., 3.

**(iii) The Multi-Client Derived Level of Assurance Aggregation (MCDLoA) Method**
If the session has several clients forming a chain (e.g., when proxies are used), then the weakest-link approach applies. In other words, the $Agg - DLoA_{session}$ is equal to the lowest link LoA value in that chain.

$$Agg - DLoA_{session} = MIN(Agg - DLoA_1, Agg - DLoA_2,$$
$$\ldots\ldots\ldots\ldots\ldots Agg - DLoA_n) \quad (4)$$

An example is when a client authenticates itself to device A (i.e., link-1), then issues a proxy to the same device to perform a task on its behalf on another device (e.g., device B). In order to do this, A needs to authenticate itself to B (i.e., link-2). If the Agg-DLoA value for the two authentication instances are 1, 2, respectively, then the $Agg - DLoA_{session}$ should be the minimum of the two values, i.e., 1.

## V. DESIGN PRELIMINARIES
### A. SYSTEM MODEL
The system model of an SHome consists of:
- Clients: human users or devices, requesting to access a target device or resources hosted by the device.
- Target devices: SHome devices used to provide a service, e.g., access to a resource.
- Services providers: human users or devices, responsible for maintaining the operation of the SHome through providing a number of services such as software updates.
- Home gateway: a coordination device that deals with resource interconnection and interoperability.

### B. THREAT MODEL
Based on the threat analysis conducted in paper [13], the threat model used in our framework is as follows.
- Internal entities are semi-trusted and curious. They often follow rules, since they may be under surveillance as they are located inside the SHome, but they may try to gain access to restricted resources or services in the SHome.
- External entities are untrustworthy. They may try to impersonate legitimate entities or intercept communications to launch a number of attacks (e.g., replay and Denial of Service (DoS) attacks) to gain access to the SHome or disrupt its availability.
- Service providers are suspicious and curious. They may track SHome entities without their consent or intercept and modify communicated data to gain access to restricted information or launch other types of attacks, e.g., DoS attacks, against their competitors.

### C. ASSUMPTIONS
(A1) Devices are classified into three groups based on their RLoA, where group 1 (G1) represents a group with the lowest required assurance level and G3 represents the group with the highest required assurance level. This assumption is made based on the fact that there is a de-facto standard for the definitions and use of the LoA for Government information systems by the National Institute of Standards and Technology (NIST), i.e., NIST 800-63-3 [37], and the European Union Agency for Cybersecurity (ENISA), i.e., the European Union regulation on electronic identification and trust services (eIDAS) [38]. Although NIST 800-63-3 and eIDAS may have different requirements for each assurance level or may compute the level of assurance differently (e.g., NIST breaks down the assurance level into independent levels to address identity proofing process, authentication process, and assertions), they both use three levels of assurance. This assumption is intended to make our solution compatible with the de-facto standard. However, the use of the three-level/group assumption does not affect the generality of our designed solution.
(A2) Each device has two symmetric keys, $K_{Di}$ and $K_{Gi}$. $K_{Di}$ is used to authenticate a device to the authentication server, whereas $K_{Gi}$ is used to verify group access credentials issued by the authentication server, and for further communication within a group of devices.
(A3) It is hard to successfully tamper with devices.

### D. NOTATIONS
The notations used in the description of the M2I protocols and Kerberos are summarised in Table 3.

### E. REQUIREMENTS
Based on the threat analysis conducted on an SHome environment in paper [13], we specify a set of requirements to secure entity authentication in IoT applications. The requirements are as follows.
- **Entity authentication** verifies the identity of a sender to a receiver and vice versa. To prevent unauthorized access and impersonation attacks, mutual authentication should be achieved during the authentication process.

AlJanah *et al.*: The M2I Authentication Framework for IoT ApplicationsTABLE 3: Notations

| Symbol | Meaning |
|---|---|
| $AD_{Ci}$ | network address of client device i |
| $EK_{Di}$ | encryption using key $K_{Di}$ |
| $EnNonce$ | random number used for authentication |
| $H_i^n$ | hash chain i of length n |
| $ID_{Ci}$ | identity of client device i |
| $ID_{Di}$ | identity of target device i |
| $ID_R$ | receiver identity |
| $ID_S$ | sender identity |
| $ID_{TGS}$ | identity of ticket granting server |
| $K-Flags$ | Kerberos flags |
| $K-SGT_{TGS}$ | Kerberos service-granting ticket issued by the ticket granting server |
| $K-TGT_{AS}$ | Kerberos ticket-granting ticket issued by the authentication server |
| $K_{Ci,Dj}$ | key shared between client device i and target device j |
| $K_{Ci,TGS}$ | key shared between client device i and the ticket granting server |
| $K_{Ci}$ | long-term key of client device i |
| $K_{Di}$ | long-term key of target device i |
| $K_{Gi}$ | key shared between group Gi devices and the authentication server |
| $K_{TGS}$ | long-term key of the ticket granting server |
| $L[Msg]$ | message length is in multiples of L bits |
| $NT$ | number of target devices |
| $Options$ | Kerberos ticket attributes request |
| $Realm_{Ci}$ | domain of client device i |
| $Seq$ | starting sequence number for messages sent to a client device in Kerberos |
| $Subkey$ | session key issued by a client device in Kerberos |
| $T_{HMAC}$ | time to perform a hashed message authentication code operation |
| $T_H$ | time to perform a hash operation |
| $T_{KSD}$ | time to perform a symmetric decryption operation in Kerberos |
| $T_{KSE}$ | time to perform a symmetric encryption operation in Kerberos |
| $T_{SE}$ | time to perform a symmetric encryption/decryption operation |
| $T_{now}$ | current time |
| $Times$ | time settings in Kerberos |
| $Ts_i$ | time stamp of entity i |
| $\triangle T$ | time interval for the allowed transmission delay |
| $W_{AuthMethod_i}$ | weight of authentication method i |

- **Message freshness** assures that the message received is fresh (i.e., it has been created recently). To counter replay and DoS attacks, a receiver should be able to verify message freshness before computing a response.
- **Confidentiality** protects the secrecy of private information, such as access credentials.
- **Authorization** verifies the access rights of a sender to a receiver. To counter unauthorized access attacks, the receiver should be able to verify the sender authorization status before processing his request.
- **Availability** ensures that the operation of the proposed authentication solution is not disrupted. In other words, the solutions is resilient against known attacks, such as DoS attacks.

In addition to the security requirements, the following functional and performance requirements are specified.

(i) **Functional Requirements**

(F1) The solution should support multi-level authentication.
(F2) The solution should support LoA based authentication decision making.
(F3) The solution should facilitate interaction based authentication, where devices are authenticated according to their mode of interaction.

(ii) **Performance Requirements**

(P1) The communication and computational costs of the protocols should be as low as possible.
(P2) The authentication delays incurred should be as low as possible.

### F. PERFORMANCE METRICS
The metrics used to evaluate protocol performance are communication and computational costs.

- **Communication Costs** are evaluated in terms of the number and length of protocol messages exchanged between entities during an authentication instance.
- **Computation Costs** are evaluated in terms of the number of cryptographic operations performed and the types of cryptographic algorithms used to perform them during an authentication instance.

### G. PROTOCOL ANALYSIS AND EVALUATION
The M2I protocols are evaluated in terms of security and performance using a number of methods as shown in Figure 1.

(i) **Methods for evaluating protocol correctness**

The following security analysis methods have been adopted to analyse the correctness of our protocols.
- Informal analysis against the specified security requirements and identified threats.
- Formal verification using the Automated Validation of Internet Security Protocols and Applications (AVISPA) verification tool.

(ii) **Methods for addressing FQ3 (What is the effectiveness of the approach?)**

- Complexity (work factor) analysis has been used to assess the computational costs required to compromise each authentication method/factor used in the protocols using brute force attacks.

(iii) **Methods for addressing FQ4 (What are the costs incurred in adopting the approach?)**

Two methods have been used to measure the cost of the multi-level authentication approach. These methods are as follows.
- Theoretical evaluation to analyze the communication and computational costs of the protocols.
- Experimental evaluation to assess the protocol cryptographic computational cost, protocol total computational cost, and authentication delays incurred during authentication.

(iv) **Additional Assumptions**
The following assumptions are used in the evaluation.
- An identifier and timestamp are each 32-bit long [39].
- A random nonce is 128-bit long [40].
- The symmetric-key cipher used is the AES-128, so the key length is 128 bits. The length of the output is in multiples of 128 bits [41].
- The hash functions used are SHA-256 and HMAC-SHA256 algorithms. Therefore, the length of any hashed value is 256 bits [42].

It is worth-noting that, as the total length of the header fields is identical in all messages, as discussed in Section VI-C, the header fields of the messages are not presented during the evaluation and performance comparison of the protocols.

## VI. THE MULTI-FACTOR MULTI-LEVEL AND INTERACTION BASED (M2I) AUTHENTICATION FRAMEWORK

### A. ARCHITECTURE

The M2I authentication architecture has two functional blocks: (i) Authentication Coordination Block, and (ii) Authorization Block as shown in Figure 2. In block (i), four functional components have been proposed to coordinate the authentication process. These components are as follows.

- Coordinator: to facilitate internal (i.e., within the block) and external (i.e., cross-block) communications.
- Negotiation: to enable flexible authentication based on a number of attributes such the RLoA value and the type of interaction (e.g., user-to-system/device or device-to-device interactions), where a client chooses how to be authenticated from a pool of authentication methods.
- Level of Assurance Derivation Module (LoADM): to derive the LoA of an authentication instance.
- Level of Assurance Aggregation Module (LoAAM): to aggregate the LoA of an authentication session.

Although the main purpose of the M2I framework is to authenticate clients, it is important to address authorization to strengthen the system against attacks, such as unauthorized access attacks, and reduce unnecessary cost [13]. This is done in block (ii) where two functional components have been proposed to verify the authorization level of the client and issue access credentials. The components are as follows.

- Access Control Function (ACF): to facilitate external communications and issue access credentials.
- Access Decision Function (ADF): to verify the authorization level of a client.

### B. THE AUTHENTICATION PROCESS

The authentication process is shown in Figure 2 and explained below.

Step 1: At the start of the authentication process, a client sends a request to the authentication coordination block to obtain a credential to access a target device.

Step 2: Upon the receipt of the request, the coordinator forwards it to the negotiation component. Depending on the type of interaction, the component replies with a list of authentication methods to obtain the RLoA to access the target device.

Step 3: Upon the receipt of the list, the client chooses suitable methods to verify its identity to an authentication server (AS) or an identity provider (IdP).

Step 4: If verified, the coordinator sends the authentication results to the LoADM component. The component derives the $Agg - DLoA_{instance}$ and sends it to the coordinator.

Step 5: If the $Agg - DLoA_{instance} >= RLoA$ to access the target device, the coordinator forwards the client request to the authorization block. Otherwise, step 2 and step 3 are repeated. Then, the authentication results are sent to the LoAAM component. The component derives the $Agg - DLoA_{session}$ and sends it to the coordinator. If $Agg - DLoA_{session} >= RLoA$, the coordinator forwards the request to the authorization block. Otherwise, the coordinator may choose to repeat this step (i.e., step 5) or terminate the authentication process.

Step 6: Upon the receipt of the request, the ACF component forwards it to the ADF component to verify the client authorization level. If authorized, the ACF issues an access credential and sends it to the client.

Step 7: The client uses its access credential to verify its identity and gain access to the target device.

### C. MESSAGE FORMAT

The message format for the M2I protocols is shown in Figure 3. Each message consists of a header field and a payload field. The header field has a fixed length and consists of five headers: Protocol Type ($ProT$), Message Type ($MsgT$), Sender Identity ($ID_S$), Receiver Identity ($ID_R$), and Payload Length ($PayL$). As each device may support the use of more than one protocol, the use of $ProT$ allows a sending device to indicate to the receiving device for which protocol the incoming message is for. Each protocol consists of two messages, a request ($Req$) and a response ($Rep$), which are identified by the $MsgT$ header. The $PayL$ header field is used to inform the message recipient about the length of the payload (measured in bytes) in the message.

The length of the message header is 12 bytes in total. It is in multiples of 32-bits to ensure memory alignment as many computers use a memory word of 4 bytes. The variable length of the message payload is discussed below.

### D. TICKET BASED PROTOCOLS

This section presents two ticket based protocols: (i) the Peer-to-Peer (P2P) protocol for device-to-device authentication, and (ii) the One-to-Many (O2M) protocol for device-to-multiDevice authentication. Similar to Kerberos, these protocols use tickets to authenticate their clients. However, the number and content of the tickets and protocol messages are different from those of Kerberos. This is because Kerberos uses two tickets (a ticket-granting ticket and a service-granting ticket) [43], whereas the P2P and O2M protocol use one ticket to verify the identity of a client. The Kerberos is designed for SSO (single sign-on), i.e., the scenarios where one user/client is accessing multiple different servers which do not belong to the same group, whereas these protocols are designed for one user/client to access a single device or multiple devices of the same group. Before discussing the protocols, the section introduces the tickets used to carry access credentials in the protocols.

#### 1) Tickets

A ticket is a temporary encrypted secret issued by the authentication server to enable a client to authenticate itself to a single device or multiple devices. This section describes the ticket structure, types, and potential clients.

##### a: Ticket Structure

The ticket structure has several fields that are used to identify the ticket owner, target, and properties, as shown in Figure 4. These fields are as follows.

- **Ticket-type** specifies the type of the ticket. Two types are defined: (1) Peer-to-Peer (P2P), and (2) One-to-Many (O2M) ticket. This is the only field visible to the client. Depending on the type of the ticket, the remaining fields are encrypted using the target device long-term key or group key.
- **ID**$_{Client}$ indicates the identity of the ticket owner.
- **Flags** represent the settings of the ticket. A flag is set if it has the value 1, and it is absent if it has the value 0. The flags are as follows.

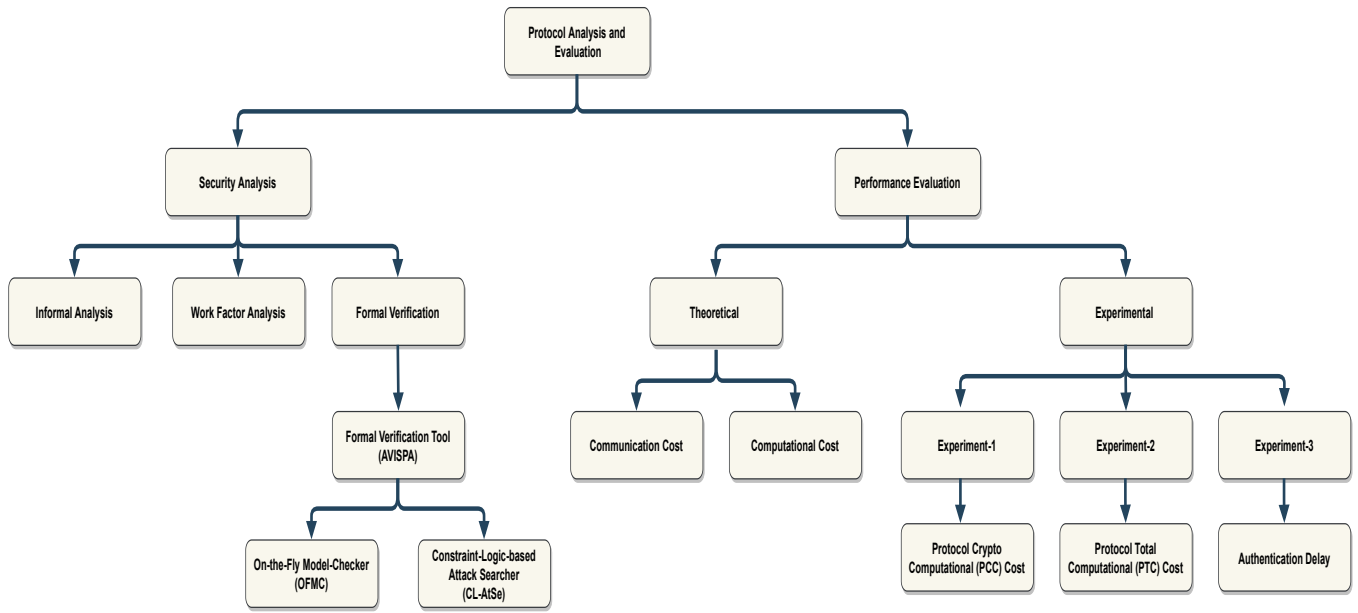

FIGURE 1: Protocol analysis and evaluation methods

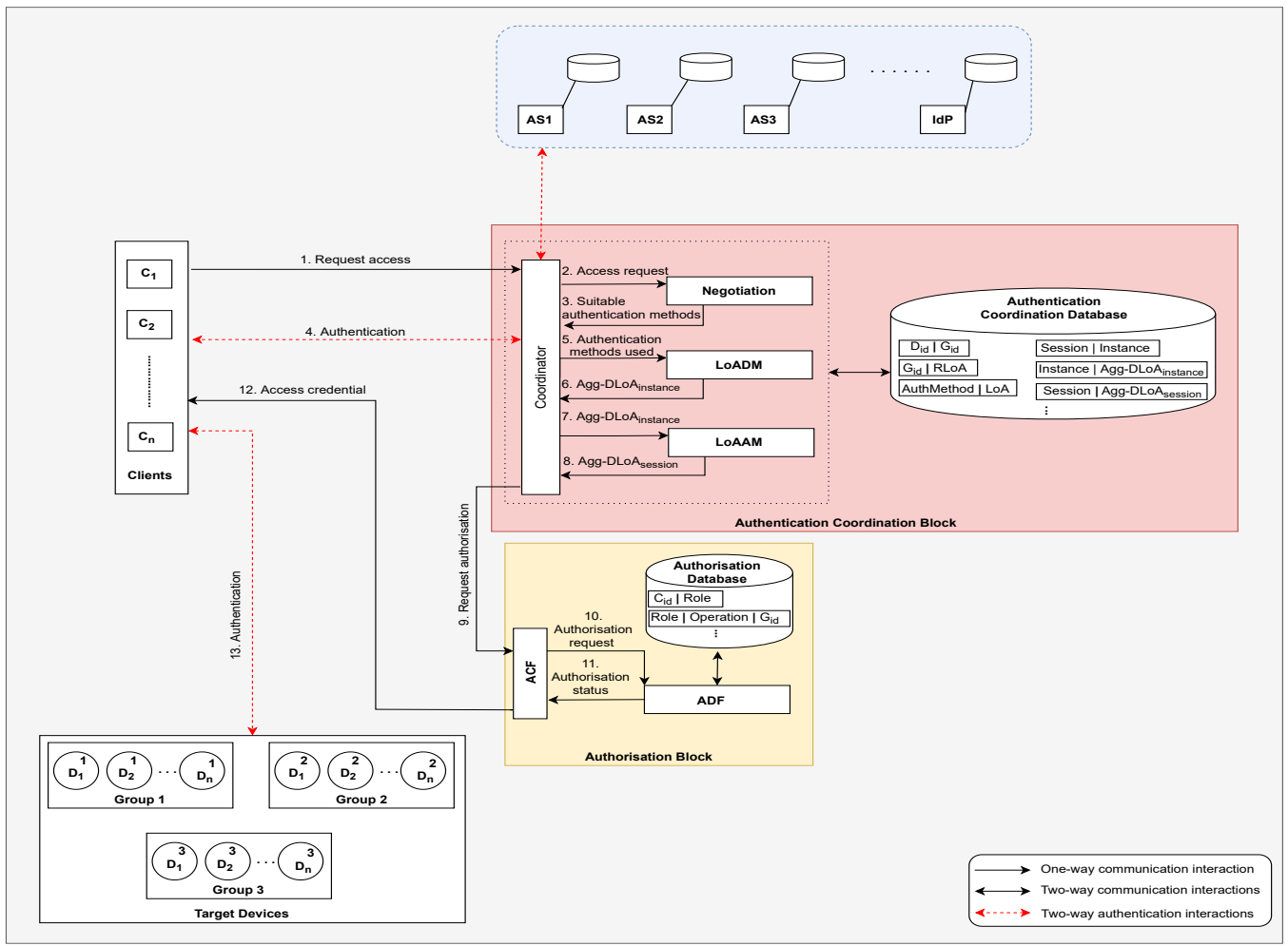

FIGURE 2: The M2I authentication architecture

| Message Length (MsgL) | Header | | | | | Payload |
|---|---|---|---|---|---|---|
| | ProT | MsgT | $ID_S$ | $ID_R$ | PayL | Payload data |
| Bit-length of each field (bits) | 4 | 1 | 32 | 32 | 27 | Variable |
| Total length (bytes) | 12 | | | | | |

FIGURE 3: Message format

**Ticket-type**

Ticket-info — $EK_{\text{128-bit key}}$
- $ID_{Client}$
- Flags
  - Renewable
  - Forwardable
  - Reusable
  - Proxiable
  - Proxy
- Session-key
- Authentication-time
- Start-time *<Optional>*
- End-time
- Renewal-deadline *<Optional>*
- LoA
- Restrictions *<Optional>*
- EnNonce

FIGURE 4: Ticket structure

- **Renewable** indicates if the ticket can be renewed. If set, the client can use the ticket to request a new ticket.
- **Forwardable** allows ticket forwarding. If set, the client can use the ticket to request a new ticket to access a different device.
- **Reusable** indicates if the ticket can be used more than once. If set, the client can use the ticket to re-authenticate itself once the authentication session has expired.
- **Proxiable** allows the client to pass a ticket to a device to perform a task on its behalf. If set, the device can use the ticket to ask the authentication server for a new ticket to perform the requested task.
- **Proxy** indicates if the ticket is a proxy passed by a different client.

- **Session-key** holds a temporary symmetric key.
- **Authentication-time** specifies the time of authentication.
- **Start-time** indicates when the ticket can be used. If it has no value, then the authentication-time is the start-time.
- **End-time** indicates the time when the ticket can no longer be used.
- **Renewal-deadline** indicates the time when the ticket can no longer be renewed.
- **LoA** represents the level of assurance encapsulated in the ticket.
- **Restrictions** field contains authorization data used to limit access privileges (e.g., a client issuing a proxy that is valid for a specific operation).
- **EnNonce** field contains a random number set by the client.

Timing values, e.g., authentication-time, are expressed using GeneralizedTime type. The GeneralizedTime syntax is *YYYYMMDDHHMMSSZ*, where four digits are used for the year, two for the month, two for the day, two for the hour, two for the minutes, and two for the seconds, followed by the letter Z to indicate the use of the Coordinated Universal Time (UTC) [44].

b: Ticket Types

Depending on the number of target devices a client intends to access, we define two types of tickets.

1) **Peer-to-Peer (P2P) Ticket**
   The P2P-Ticket is used for device-to-device authentication. It is an access credential issued by the authentication server to enable the client device to authenticate itself to a single target device. The P2P-Ticket is encrypted using the target device long-term key $K_{Di}$. The content of the ticket is as follows.

   $P2P - Ticket :=< P2P, EK_{Di}[ID_{Ci}, Flags,$
   $Session - key, Authentication - time,$
   $Start - time, End - time, Renewal - deadline,$
   $LoA, Restrictions, EnNonce] >$

2) **One-to-Many (O2M) Ticket**
   The O2M-Ticket is used for device-to-multiDevice authentication. It is an access credential issued by the authentication server to enable the client device to authenticate itself to a group of devices. The O2M-Ticket is encrypted using the target device group key $K_{Gi}$. The content of the ticket is as follows.

   $O2M - Ticket :=< O2M, EK_{Gi}[ID_{Ci}, Flags,$
   $Session - key, Authentication - time,$
   $Start - time, End - time, Renewal - deadline,$
   $LoA, Restrictions, EnNonce] >$

c: Ticket Requestors

The device class dictates its ability to request a ticket and the properties of the ticket. For instance, a lower class device, e.g., $C_1$ device, should not be able to obtain a ticket that is valid for a long period of time. This is because the higher the class, the more security services the device can offer [45]. Table 4 presents an example of who can request which types of tickets.

d: Communication Cost

The authentication tokens used in the P2P and O2M protocol follow the same structure and format of an M2I ticket. The fields and the bit-length of each field in an M2I ticket is shown in Figure 5. From the figure, it can be seen that the ticket consists of two main fields: a Ticket-type field and a Ticket-info field. The Ticket-type field is 1-bit long to indicate one of the two types of tickets used in our protocols. The Ticket-info field consists of a further number of fields and the total length for this field is 512 bits.

| M2I ticket | | Bit-length of each field (bits) | Total length (bits) |
|---|---|---|---|
| Ticket-type | | 1 | |
| Ticket-info | $ID_{Client}$ | 32 | 513 |
| | Flags | 8 | |
| | Renewable | | |
| | Forwardable | | |
| | Reusable | | |
| | Proxiable | | |
| | Proxy | | |
| $EK_{128\text{-bit key}}$ | Session-key | 128 | |
| | Authentication-time | 32 | |
| | Start-time | 32 | |
| | End-time | 32 | |
| | Renewal-deadline | 32 | |
| | LoA | 4 | |
| | Restrictions | 4 | |
| | EnNonce | 128 | |

FIGURE 5: Communication cost of the M2I ticket

2) Peer-to-Peer (P2P) Protocol for Device-to-Device Authentication

The P2P protocol is designed for device-to-device authentication. The protocol uses a token issued by the authentication server, i.e., P2P-Ticket, and an authenticator generated by a client device to verify the client identity to a target device and achieve mutual authentication, as shown in Figure 6. This section details the P2P protocol messages and operation description.

a: Protocol Messages

The protocol consists of five messages for authentication and two for re-authentication, as shown in Figure 6 and Figure 7.

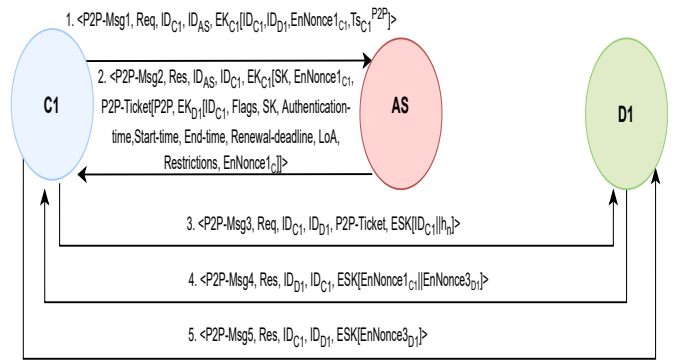

FIGURE 6: P2P message exchange diagram

b: Operation Description

The P2P protocol operations are explained below.

**Step S1-P2P:** At the start of the protocol, client C1 generates a fresh EnNonce and timestamp. Then, it constructs and sends Msg1 to the AS. Once Msg1 is sent, C1 starts a timer and await for a timeout. If no response is received upon the expiry of this timeout, it will either resend the message or terminate the protocol execution.

**Step S2-P2P:** Upon the receipt of Msg1, the AS decrypts $EK_{C1}[ID_{C1}, ID_{D1}, EnNonce1_{C1}, Ts_{C1}^{P2P}]$ using $K_{C1}$ to verify the freshness of $Ts_{C1}^{P2P}$ using the TS-Veri algorithm. If fresh, the AS verifies C1 identity using the ID-Veri algorithm. Then, it generates a session key and a P2P-Ticket to construct and send Msg2.

**Step S3-P2P:** Upon the receipt of Msg2, C1 decrypts $EK_{C1}[SK, EnNonce1_{C1}, P2P-Ticket]$ using $K_{C1}$ to verify $EnNonce1_{C1}$ using the EN-Veri algorithm. Then, it generates $EnNonce2_{C1}$ and uses it to generate a fresh authenticator ($ESK[ID_{C1}||EnNonce2_{C1}]$) if it is a non-reusable ticket. However, if the ticket is reusable, C1 uses $EnNonce2_{C1}$ as seed to compute a hash chain of length n ($H_i^n$), as shown in Figure 8, where n is the number of times the client intends to use the ticket to authenticate itself. Once $H_i^n$ is computed, C1 uses the last link in the chain, i.e., $h_n$, to generate the fresh authenticator ($ESK[ID_{C1}||h_n]$). Then, it constructs Msg3 and sends it to the target device D1 to request access.

$H_i^n(EnNonce) = H(EnNonce), H(H(EnNonce)), \ldots\ldots H(h_{n-1})$

$h_0 \quad h_1 \quad h_n$

FIGURE 8: Hash chain

**Step S4-P2P:** Upon the receipt of Msg3, D1 decrypts the P2P-Ticket using $K_{D1}$ to verify it using the TI-Veri algorithm. Then, it decrypts the attached authenticator using the SK obtained from the ticket to verify the ownership of the ticket using the ID-Veri algorithm. If the ticket is reusable, D1 saves the hashed EnNonce value (i.e., $h_n$) for future authentication requests. After that, it generates $EnNonce3_{D1}$ to construct and send Msg4.

TABLE 4: Ticket requestors

| Device class | Device capabilities | | Ticket type | |
|---|---|---|---|---|
| | Data size $(KiB)$ | Code size $(KiB)$ | P2P-Ticket | O2M-Ticket |
| $C_0$ | <10 | <100 | No | No |
| $C_1$ | ~10 | ~100 | Yes | No |
| $C_2$ | ~50 | ~250 | Yes | Yes |
| $C_{2+}$ | >50 | >250 | Yes | Yes |

KiB = 1024 bytes

| Entities | Protocol messages | Items |
|---|---|---|
| $C_1 \to AS$ | Msg1:= | $< P2P - Msg1, Req, ID_{C1}, ID_{AS}, EK_{C1}[ID_{C1}, ID_{D1},$ $EnNonce1_{C1}, Ts^{P2P}_{C1}] >$ |
| $AS \to C_1$ | Msg2:= | $< P2P - Msg2, Res, ID_{AS}, ID_{C1}, EK_{C1}[SK, EnNonce1_{C1},$ $P2P - Ticket] >$ $P2P - Ticket = [P2P, EK_{D1}[ID_{C1}, Flags, SK,$ $Authentication - time, Start - time, End - time,$ $Renewal - deadline, LoA, Restrictions, EnNonce1_{C1}]$ |
| $C_1 \to D_1$ | Msg3:= | $< P2P - Msg3, Req, ID_{C1}, ID_{D1}, P2P - Ticket,$ $ESK[ID_{C1}||h_n] >$ |
| $D_1 \to C_1$ | Msg4:= | $< P2P - Msg4, Res, ID_{D1}, ID_{C1}, ESK[EnNonce1_{C1}||$ $EnNonce3_{D1}] >$ |
| $C_1 \to D_1$ | Msg5:= | $< P2P - Msg5, Res, ID_{C1}, ID_{D1}, ESK[EnNonce3_{D1}] >$ |
| | For re-authentication | |
| $C_1 \to D_1$ | Msg6:= | $< P2P - reauth - Msg1, Req, ID_{C1}, ID_{D1}, P2P - Ticket,$ $ESK[ID_{C1}||h_{n-1}] >$ |
| $D_1 \to C_1$ | Msg7:= | $< P2P - reauth - Msg2, Res, ID_{D1}, ID_{C1}, ESK[h_{n-1}||$ $EnNonce4_{D1}] >$ |

FIGURE 7: P2P protocol messages

**Step S5-P2P:** Upon the receipt of Msg4, C1 decrypts $ESK[EnNonce1_{C1}||EnNonce3_{D1}]$ using the SK to verify $EnNonce1_{C1}$ using the EN-Veri algorithm. Then, it constructs Msg5 and sends it to D1 to verify its own identity and achieve mutual authentication.

**Step S6-P2P:** Upon the receipt of Msg5, D1 decrypts $ESK[EnNonce3_{D1}]$ using the SK to verify $EnNonce3_{D1}$ using the EN-Veri algorithm. If verified, C1 and D1 achieve mutual authentication and the protocol is terminated.

For subsequent device access requests using the same credential, i.e., the same ticket, only two messages are needed to achieve mutual authentication, as shown in Figure 9.

**Step S1-P2P2:** C1 generates a fresh authenticator using the link that precedes the last used link in $H_i^n$, i.e., $h_{n-1}$. Then, it constructs and sends Msg6.

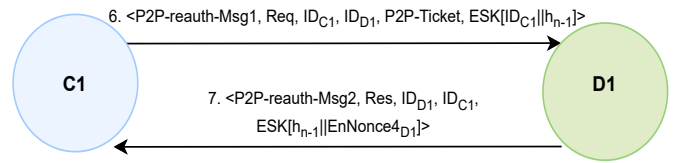

FIGURE 9: P2P re-authentication message exchange diagram

**Step S2-P2P2:** Upon the receipt of Msg6, D1 decrypts the P2P-Ticket using $K_{D1}$ to verify it using the TI-Veri algorithm. Then, it decrypts the attached authenticator using the SK to verify it using the ID-Veri and HC-Veri algorithms. After that, it generates $EnNonce4_{D1}$ to construct and send Msg7 to verify its own identity and achieve mutual authentication.

**Step S3-P2P2:** Upon the receipt of Msg7, C1 decrypts $ESK[h_{n-1}||EnNonce4_{D1}]$ using the SK to verify $h_{n-1}$ using the EN-Veri algorithm. If verified, C1 and D1 achieve mutual authentication and the protocol is terminated.

If any of the verifications is negative, the protocol is immediately terminated.

#### c: Performance Evaluation
**(i) Communication Cost**
The total communication cost of one execution of the P2P protocol is ($2433 \times$ the number of target devices (NT)) bits, as shown in Table 5. For example, if a client device authenticates itself to three target devices using the P2P protocol, the communication cost incurred would be ($2433 \times 3$) bits = 913 bytes.

The total communication cost for re-authentication using the P2P protocol is ($1281 \times NT$) bits, as presented in Table 6. For example, if a client device re-authenticates itself to three target devices using the protocol after its initial authentication has expired, the communication cost incurred would be ($1281 \times 3$) bits = 481 bytes.

**(ii) Computation Cost**
The total computation cost of one execution of the P2P protocol is $NT(12T_{SE} + T_H)$ ms, as shown in Table 7. For example, if a client device authenticates itself to three target devices using the protocol, the computation cost incurred would be $3(12T_{SE} + T_H) = (36T_{SE} + 3T_H)$ ms.

The total computation cost to re-authenticate a client device is $NT(5T_{SE} + T_H)$ ms. For example, if a client device re-authenticates itself to 3 target devices after its initial authentication has expired, the computation cost would be $3(5T_{SE} + T_H) = (15T_{SE} + 3T_H)$ ms.

### 3) One-to-Many (O2M) Protocol for Device-to-multiDevice Authentication

To recognise the fact that (i) in an IoT environment, most devices can be grouped due to reasons such as they are physically in a close proximity and/or perform the same function, e.g., devices in the same room, or the light switches in the whole house, (ii) the devices in the same group may have similar security requirement, (iii) there are cases where an interaction is to a group of devices, and (iv) most IoT devices are resource-constrained, it is desirable to have an authentication solution that takes into account all of these characteristics and introduce as low cost as possible. Our proposed solution, the O2M protocol, is designed to allow a client device to authenticate itself to a group of devices which perform a similar function or with a similar security requirement using the same access credentials. The protocol uses a token, i.e., O2M-ticket, issued by the authentication server, and an authenticator generated by the client to verify the client identity to a group of target devices and achieve mutual authentication.

#### a: Protocol Messages
The protocol consists of $2 + (3 \times NT)$ messages, where $NT$ is the number of target devices in the group, as depicted in Figure 10.

#### b: Operation Description
The O2M protocol performs the same operations as the P2P protocol with a few differences. The differences are as follows.

1) The O2M protocol uses an O2M-Ticket that could be verified by a group of devices as it is encrypted using their group key $K_{Gi}$. On the other hand, the P2P protocol uses a P2P-Ticket that could only be verified by one device as it is encrypted using the target device long-term key $K_{Di}$.
2) The O2M protocol repeats the last four steps (i.e., S3 to S6) in the P2P protocol operation description $NT$ times. This is to achieve mutual authentication between the client and all target devices involved.

Similar to the P2P protocol, the protocol is immediately terminated if any of the verifications is negative.

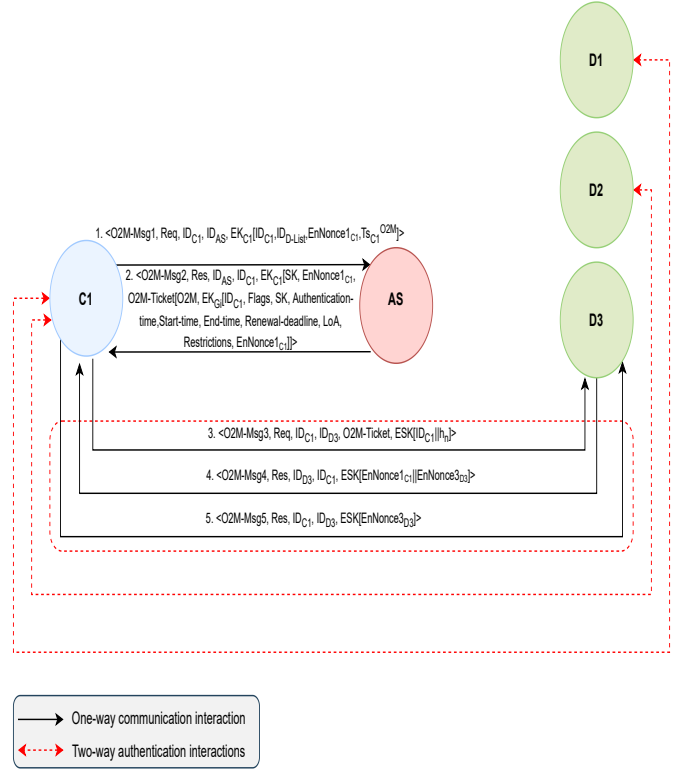

FIGURE 10: O2M message exchange diagram

#### c: Performance Evaluation
**(i) Communication Cost**
The total communication cost of one execution of the O2M protocol is $((1281 \times NT) + 128[(32 \times NT) + 192] + 896)$ bits as presented in Table 8. The three messages, Msg3, Msg4 and Msg5, are repeated for each of the target devices involved. For example, if a client device authenticates itself to three target devices using the protocol, the communication cost incurred would be $((1281 \times 3) + 128[(32 \times 3) + 192] + 896)$ bits = 641 bytes.

The total communication cost for re-authentication using the protocol is identical to the P2P protocol. This is because the number and length of re-authentication messages are the same in both protocols.

**(ii) Computation Cost**
The total computation cost of one execution of the O2M protocol is $(5T_{SE} + NT(7T_{SE} + T_H))$ ms, as described in Table 9. For example, if a client device authenticates itself to three target devices using the protocol, the computation cost incurred would be $5T_{SE} + 3(7T_{SE} + T_H) = (26T_{SE} + 3T_H)$ ms.

The total computation cost for re-authentication using the protocol is identical to the P2P protocol. This is because both protocols use the same number of cryptographic operations and type of cryptographic algorithms to re-authenticate their client devices.

TABLE 5: Communication cost of the P2P protocol

| Entities | Protocol messages | Items | Total length ($bits$) |
|---|---|---|---|
| Client device | Msg1 | $EK_{C1}[ID_{C1}, ID_{D1}, EnNonce1_{C1}, Ts^{P2P}_{C1}]$ | 256 |
| AS | Msg2 | $EK_{C1}[SK, EnNonce1_{C1}, P2P-Ticket]$ | 896 |
| Client device | Msg3 | $P2P-Ticket, ESK[ID_{C1}||h_n]$ | 897 |
| Target device | Msg4 | $ESK[EnNonce1_{C1}||EnNonce3_{D1}]$ | 256 |
| Client device | Msg5 | $ESK[EnNonce3_{D1}]$ | 128 |
| The total length per protocol execution | | | $2433 \times NT$ |

TABLE 6: Communication cost for re-authentication using the P2P protocol

| Entities | Protocol messages | Items | Total length ($bits$) |
|---|---|---|---|
| Client device | Msg6 | $P2P-Ticket, ESK[ID_{C1}||h_{n-1}]$ | 897 |
| Target device | Msg7 | $ESK[h_{n-1}||EnNonce4_{D1}]$ | 384 |
| The total length per protocol execution | | | $1281 \times NT$ |

TABLE 7: Computation cost of the P2P protocol

| Protocols | Entities | | | Total cost ($ms$) |
|---|---|---|---|---|
| | Client device | AS | Target device | |
| P2P | $5T_{SE} + T_H$ | $3T_{SE}$ | $4T_{SE}$ | $NT(12T_{SE} + T_H)$ |
| P2P re-authentication | $2T_{SE}$ | – | $3T_{SE} + T_H$ | $NT(5T_{SE} + T_H)$ |

| Entities | Protocol messages | Items |
|---|---|---|
| $C_1 \to AS$ | Msg1:= | $< O2M-Msg1, Req, ID_{C1}, ID_{AS}, EK_{C1}[ID_{C1}, ID_{D-List}, EnNonce1_{C1}, Ts^{O2M}_{C1}] >$ |
| $AS \to C_1$ | Msg2:= | $< O2M-Msg2, Res, ID_{AS}, ID_{C1}, EK_{C1}[SK, EnNonce1_{C1}, O2M-Ticket] >$ <br> $O2M-Ticket = [O2M, EK_{G_i}[ID_{C1}, Flags, SK,$ <br> $Authentication-time, Start-time, End-time,$ <br> $Renewal-deadline, LoA, Restrictions, EnNonce1_{C1}]$ |
| $C_1 \to D_3$ | Msg3:= | $< O2M-Msg3, Req, ID_{C1}, ID_{D3}, O2M-Ticket,$ <br> $ESK[ID_{C1}||h_n] >$ |
| $D_3 \to C_1$ | Msg4:= | $< O2M-Msg4, Res, ID_{D3}, ID_{C1}, ESK[EnNonce1_{C1}||$ <br> $EnNonce3_{D3}] >$ |
| $C_1 \to D_3$ | Msg5:= | $< O2M-Msg5, Res, ID_{C1}, ID_{D3}, ESK[EnNonce3_{D3}] >$ |
| | For re-authentication | |
| $C_1 \to D_3$ | Msg6:= | $< O2M-reauth-Msg1, Req, ID_{C1}, ID_{D3}, O2M-Ticket,$ <br> $ESK[ID_{C1}||h_{n-1}] >$ |
| $D_3 \to C_1$ | Msg7:= | $< O2M-reauth-Msg2, Res, ID_{D3}, ID_{C1}, ESK[h_{n-1}||$ <br> $EnNonce4_{D3}] >$ |

FIGURE 11: O2M protocol messages

AlJanah *et al.*: The M2I Authentication Framework for IoT ApplicationsTABLE 8: Communication cost of the O2M protocol

| Entities | Protocol messages | Items | Total length ($bits$) |
|---|---|---|---|
| Client device | Msg1 | $EK_{C1}[ID_{C1}, ID_{D-List},$ $EnNonce1_{C1}, Ts^{O2M}_{C1}]$ | $128[(32 \times NT) + 192]$ |
| AS | Msg2 | $EK_{C1}[SK, EnNonce1_{C1},$ $O2M - Ticket]$ | 896 |
| Client device | Msg3 | $O2M - Ticket,$ $ESK[ID_{C1}||h_n]$ | 897 |
| Target device | Msg4 | $ESK[EnNonce1_{C1}||$ $EnNonce3_{D3}]$ | 256 |
| Client device | Msg5 | $ESK[EnNonce3_{D3}]$ | 128 |
| The total length per protocol execution | | | $(1281 \times NT)+$ $128[(32 \times NT) + 192] + 896$ |

TABLE 9: Computation cost of the O2M protocol

| Protocols | Entities | | | Total cost ($ms$) |
|---|---|---|---|---|
| | Client device | AS | Target device | |
| O2M | $2T_{SE}+$ $NT(3T_{SE} + T_H)$ | $3T_{SE}$ | $4T_{SE} \times NT$ | $5T_{SE}+$ $NT(7T_{SE} + T_H)$ |
| O2M re-authentication | $2T_{SE}$ | – | $3T_{SE} + T_H$ | $NT(5T_{SE} + T_H)$ |

## VII. SECURITY ANALYSIS

This section presents informal analysis, work factor analysis, and formal verification of the protocols.

### A. INFORMAL ANALYSIS

The M2I protocols are informally analysed with respect to the security requirements and identified threats which may be countered by an entity authentication service.

#### 1) Requirements analysis

A summary of the security requirements analysis of the P2P and O2M protocol against the state-of-the-art IoT authentication solutions, discussed in detail in paper [13], is presented in Table 10. Although the P2P and O2M protocol, respectively, achieve mutual entity authentication for device-to-device and device-to-multiDevice modes of interactions, the table shows that entity authentication is partially supported in the protocols. This is because the protocols do not address all modes of interactions (i.e., user-to-device, device-to-device, device-to-multiDevice, and multiDevice-to-device interactions [13]). The work on the M2I framework is still in progress to address the remaining modes of interactions.

##### a: Entity Authentication
The M2I protocols use the challenge-response mechanism where fresh random numbers, e.g., $EnNonce$, are generated by the requestor and responder to achieve mutual authentication. Receiving the wrong response will lead to instant protocol termination.

##### b: Message Freshness
Timestamps and freshly generated random numbers are common freshness identifiers. Both methods have been used in the M2I protocols to verify the freshness of the exchanged messages.

##### c: Confidentiality
In the M2I protocols, all secret authentication parameters, e.g., tickets and authenticators, are never transmitted in plaintext. They are encrypted using a symmetric cryptosystem. Provided that the key length is sufficiently large, e.g., 128 bits for the AES algorithm, it would be computationally difficult for an adversary to decrypt any of the intercepted messages.

##### d: Authorization
In the P2P and O2M protocol, the authentication server checks the client's authorization status before issuing a ticket. Once issued, authorization information can be found in the ticket itself under the restrictions field.

##### e: Availability
The M2I protocols are designed to avoid bottlenecks and be resilient to DoS attacks. As mentioned earlier, the protocols use the challenge-response mechanism. Receiving the wrong response (i.e., wrong message) at any stage of the protocol should lead to its termination, making the IoT application available to its legitimate users.

#### 2) Threat analysis

The threat analysis has been carried out under the assumption that an adversary is able to eavesdrop all messages.

##### a: Impersonation Attacks
An adversary may try to impersonate a legitimate entity to gain access, a higher privilege level, or launch attacks against an IoT environment. Potential impersonation attacks are as follows.

**(i) Client Impersonation**
In the P2P and O2M protocol, the adversary would not be able to impersonate a client to deceive the authentication server into issuing a ticket without knowing the client's long-term key, or forge a valid ticket without knowing the target device long-term key. Furthermore, even if the adversary was able to somehow capture a valid ticket, s/he would not be able to use it to deceive the target device. This is because the adversary would not be able to alter the $ID_{Client}$ field without knowing the target device long-term key, or forge a valid authenticator without knowing the session key. To obtain access, the adversary would need to have a valid ticket and a fresh authenticator.

2022　　　　　　　　　　　　　　　　　　　　　　　　　　　　　　　　　　　　　　　　　　　　　　　　　　　　　　13

**(ii) Authentication Server Impersonation**
Since the secret authentication parameters exchanged between clients and the authentication server are never transmitted in plaintext, the adversary would not be able to impersonate the server without knowing the clients' long-term keys. As a result, the proposed protocols resist authentication server impersonation attacks.

**(iii) Target Impersonation**
An adversary would not be able to impersonate a target device in the P2P protocol without knowing the device's long-term key or its group key in the O2M protocol. Therefore, the proposed protocols can resist target impersonation attacks.

### b: Eavesdropping
The M2I protocols are designed under the assumption that any message could be intercepted and hence none of the authentication parameters are transmitted in plaintext. In other words, intercepted messages are useless to the interceptor as it is computationally difficult to decipher them. Therefore, the protocols can withstand eavesdropping attacks, such as passive or active man-in-the-middle attacks.

### c: Replay
The protocols use timestamps and random numbers to counter replay attacks. The first message in each of the M2I protocols has a timestamp attached to it as its freshness identifier. In addition, subsequent protocol messages use fresh random numbers, instead of timestamps, to avoid clock desynchronisation issues. As a result, the protocols can withstand replay attacks.

### d: Denial of Service (DoS)
DoS attacks often rely on the the transmission of oversized messages and/or a large number of requests to disrupt availability, making the IoT application unavailable to its legitimate users. Owing to the characteristics of the M2I protocols, an adversary would not be able to forge a legitimate message to slow down or occupy a client, target device, or the authentication server without knowing their long-term keys. Furthermore, replayed messages are easily detected using timestamps and random numbers as discussed earlier. Therefore, the M2I protocols can resist DoS attacks.

## B. WORK FACTOR ANALYSIS
The work factor, also known as work function, is the computational cost required to compromise each authentication method/factor used in the protocols using brute force attacks. It is typically proportional to the security level of the authentication method [46]. To launch a successful attack on a method that provides n-bit security level, a computational complexity of $2^n$ is needed [47]. For example, the work factor needed to launch successful attack on AES-128 is $2^{128}$. Depending on the key-length and the cryptographic algorithms used, authentication methods may provide different security levels as shown in Table 11 [48].

To have a cryptosystem that is secure beyond 2030, the National Institute of Standards and Technology (NIST) suggests that the security level provided should not be less than 128-bit [48]. Thus, the M2I protocols use AES-128 as their symmetric cryptosystem, SHA-256 and HMAC-SHA256 as their hash and HMAC algorithms, respectively, to comply with NIST's recommendation. Table 12 shows the computational complexity needed to forge a successful device access request in each of the M2I protocols. To forge a successful device access request in the P2P protocol, the target device long term-key ($K_{Di}$) have to be compromised. Similarly, the group key of the target device ($K_{Gi}$) needs to be compromised to forge a successful device access request in the O2M protocol. As a result, their work factor is $2^{128}$.

TABLE 12: Work factor

| Protocol | Work factor |
|---|---|
| P2P | $2^{128}$ |
| O2M | $2^{128}$ |

## C. FORMAL SECURITY VERIFICATION
A number of formal verification methods, e.g., AVISPA [49], ProVerif [50], and Scyther [51], can be used to evaluate the security of a protocol. AVISPA (Automated Validation of Internet Security Protocols and Applications) tool has been chosen to validate our protocols for the following reasons. First, the AVISPA tool models protocols and their security properties in High Level Protocol Specification Language (HLPSL). HLPSL is a powerful and expressive language that provides abstraction [49]. Secondly, it integrates different verification tools, e.g., On-the-Fly Model-Checker (OFMC) and Constraint-Logic-based Attack Searcher (CL-AtSe), that use a variety of analysis techniques to verify protocol correctness [52]. Lastly, the tool has been widely used to verify authentication protocols proposed for IoT applications [12, 18, 23, 28–30].

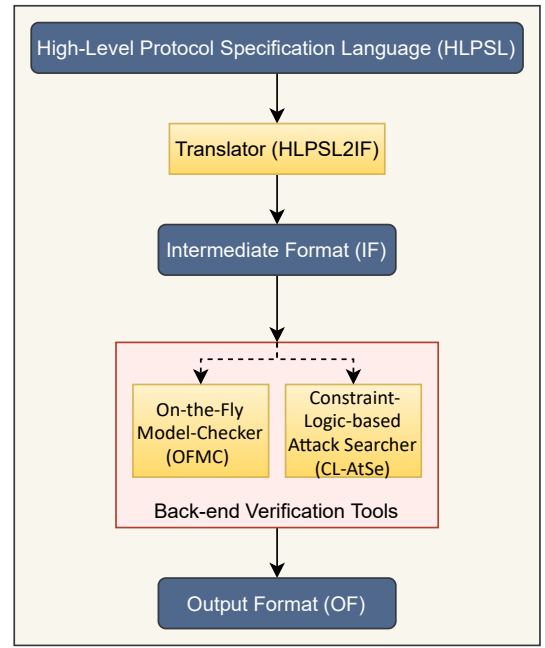

FIGURE 12: AVISPA architecture

As shown in Figure 12, the AVISPA tool starts by translating the HLPSL specification into a lower level specification, known as intermediate format (IF). The IF specification is then used as input to the back-end verification tools, e.g., OFMC and CL-AtSe. The tools apply a number of automatic analysis techniques to verify protocol correctness against specified security requirements. The results of the verification process is then displayed in a format known as the output format (OF) [49].

The M2I protocols have been formally verified in terms of entity authentication, confidentiality, and resilience against known attacks. The results of the verification are presented in Figure 13 and Figure 14. A summary of this verification is given in Table 13. The HLPSL code of the protocols is presented in Appendix B.

TABLE 10: Security requirements analysis of the P2P and O2M protocol vs IoT authentication solutions

| The State-of-the-Art | | | R1 | R2 | R3 | R4 | R5 |
|---|---|---|---|---|---|---|---|
| Non-cryptographic | | Tewari and Gupta [14] | o | x | x | x | x |
| | | Fan et al. [15] | o | o | x | x | x |
| | | Martinez and Bossuet [16] | o | x | x | x | x |
| | | Gu et al. [17] | o | x | x | x | ✓ |
| Cryptographic | Symmetric-key based | Amin et al. [18] | o | o | o | x | x |
| | | Wu et al. [19] | o | x | o | x | ✓ |
| | | Wazid et al. [20] | o | ✓ | o | ✓ | x |
| | | Fotouhi et al. [21] | o | x | x | x | ✓ |
| | | Liu et al. [22] | o | ✓ | ✓ | x | ✓ |
| | | Gope et al. [12] | o | x | x | x | x |
| | | Lara et al. [23] | o | ✓ | o | x | ✓ |
| | | Mahalat et al. [24] | o | x | x | x | x |
| | | Liang et al. [25] | o | x | x | x | x |
| | | Fan et al. [26] | o | ✓ | o | x | x |
| | | Lai et al. [27] | o | x | x | x | x |
| | | Modiri et al. [28] | o | ✓ | x | x | ✓ |
| | Asymmetric-key based | Chen et al. [29] | o | ✓ | o | x | x |
| | | Nikravan and Reza [30] | o | ✓ | ✓ | x | x |
| | | Chatterjee et al. [31] | o | o | x | x | x |
| | | Braeken [32] | o | ✓ | x | x | x |
| | | Naeem et al. [33] | o | x | o | x | x |
| | | Izza et al. [34] | o | o | o | x | x |
| | | Shen et al. [35] | o | x | ✓ | x | x |
| | | Liu et al. [36] | o | x | ✓ | x | ✓ |
| The M2I protocols | | | | | | | |
| Cryptographic | Symmetric-key based | The P2P protocol | o | ✓ | ✓ | ✓ | ✓ |
| | | The O2M protocol | o | ✓ | ✓ | ✓ | ✓ |

✓: supported ; x: not supported; o: partially supported.
R1: entity authentication; R2: message freshness; R3: confidentiality; R4: authorization; R5:availability.

TABLE 11: Security level of cryptographic algorithms

| Security level ($bits$) | Symmetric-key algorithms | Hash algorithms | HMAC algorithms |
|---|---|---|---|
| $<= 80$ | 2TDEA | SHA-1 | - |
| 112 | 3TDEA | SHA-224, SHA-512/224 | - |
| 128 | AES-128 | SHA-256, SHA-512/256 | SHA-1 |
| 192 | AES-192 | SHA-384 | SHA-224, SHA-512/224 |
| 256 | AES-256 | SHA-512 | SHA-256, SHA-512/256, SHA-384, SHA-512 |

AES: Advanced Encryption Standard; SHA: Secure Hash Algorithm; 2TDEA: Two-key Triple Data Encryption Algorithm; 3TDEA: Three-key Triple Data Encryption Algorithm.

```
SUMMARY
 SAFE
DETAILS
 BOUNDED_NUMBER_OF_SESSIONS
PROTOCOL
 /home/span/span/testsuite/results/P2P.if
GOAL
 as_specified
BACKEND
 OFMC
COMMENTS
STATISTICS
 parseTime: 0.00s
 searchTime: 0.22s
 visitedNodes: 144 nodes
 depth: 8 plies
```
(a) OFMC

```
SUMMARY
 SAFE
DETAILS
 BOUNDED_NUMBER_OF_SESSIONS
 TYPED_MODEL
PROTOCOL
 /home/span/span/testsuite/results/P2P.if
GOAL
 As Specified
BACKEND
 CL-AtSe
STATISTICS
 Analysed  : 8 states
 Reachable : 0 states
 Translation: 0.04 seconds
 Computation: 0.00 seconds
```
(b) CL-AtSe

FIGURE 13: AVISPA results of the P2P protocol

```
SUMMARY
 SAFE
DETAILS
 BOUNDED_NUMBER_OF_SESSIONS
PROTOCOL
 /home/span/span/testsuite/results/O2M.if
GOAL
 as_specified
BACKEND
 OFMC
COMMENTS
STATISTICS
 parseTime: 0.00s
 searchTime: 0.13s
 visitedNodes: 81 nodes
 depth: 8 plies
```
(a) OFMC

```
SUMMARY
 SAFE
DETAILS
 BOUNDED_NUMBER_OF_SESSIONS
 TYPED_MODEL
PROTOCOL
 /home/span/span/testsuite/results/O2M.if
GOAL
 As Specified
BACKEND
 CL-AtSe
STATISTICS
 Analysed  : 8 states
 Reachable : 0 states
 Translation: 0.03 seconds
 Computation: 0.00 seconds
```
(b) CL-AtSe

FIGURE 14: AVISPA results of the O2M protocol

TABLE 13: A summary of the formal verification results

| Protocol | AVISPA | |
|---|---|---|
| | OFMC | CL-AtSe |
| P2P | ✓ | ✓ |
| O2M | ✓ | ✓ |

✓: Safe; x: Unsafe.

## VIII. THE EXPERIMENTS

The experiments are carried out to experimentally evaluate the computational costs and authentication delays incurred in each of the M2I protocols. This section gives the experiment design and set-ups, and discusses the results.

### A. EXPERIMENT DESIGN

The experiment design covers the selections of the programming language and cryptographic algorithms used to implement the protocols, machine set-up and specifications, and definitions of performance metrics.

#### 1) Programming Language

The programming language used to implement the M2I protocols is Python 3.7. Python was chosen because it supports a cryptography package, known as PyCryptodome. PyCryptodome provides the implementation of several cryptographic primitives and key management services used in our protocols, including a secure random number generator, a collection of message digest functions, and several encryption algorithms [53].

#### 2) Cryptographic Algorithms

The cryptographic algorithms used in the implementation of our protocols are as follows.

- AES algorithm (using the Cipher Block Chaining (CBC) mode) with a key length of 128 bits is used for symmetric encryption/decryption.
- The SHA-256 and HMAC-SHA256 algorithms are used to generate hash and HMAC values, respectively.

The cryptographic algorithm used in the implementation of Kerberos version 5 is as follows.

- AES algorithm (using the Cipher Block Chaining with Ciphertext Stealing (CBC-CTS) mode) with a key length of 128 bits is used for symmetric encryption/decryption.

#### 3) Machine Set-up and Specifications

The implementation of the M2I protocols has been carried out under two experiment set-ups: (i) 1-machine set-up, and (ii) 2-machine set-up. In case (i), a single machine is used to run all the entities, i.e., the client devices, the server, and target devices. This set-up is used to measure the computational costs of the cryptographic operations, and the total computational cost introduced by all the operations (i.e., cryptographic and non-cryptographic operations) in each of the protocols. In case (ii), two machines are used. The first machine is used to run the client device while the second machine is used to run the server, and target devices. This set-up is used to evaluate authentication delays introduced by the protocols. As authentication requests and responses are typically sent by different machines, using the 2-machine set-up to measure authentication delays is more adequate than using the 1-machine set-up. The specifications of the two machines used are as follows:

- **Machine-1 (M1)** is a laptop computer running Windows 10 (64-bit operating system) with a 1.60 GHz Intel Core i5-8265U CPU and 12 GB of RAM memory.
- **Machine-2 (M2)** is a laptop computer running Windows 10 (64-bit operating system) with a 1.80 GHz Intel Core i7-10510U CPU and 16 GB of RAM memory.

#### 4) Performance Metrics

The metrics used to evaluate our protocol performance are protocol crypto computational cost, protocol total computational cost, and authentication delay.

- **Protocol Crypto Computational (PCC) cost** of a protocol is defined as the time taken to perform all the cryptographic operations during the execution of a protocol.
- **Protocol Total Computational (PTC) cost** of a protocol is defined as the time taken to perform all the operations during the execution of a protocol. This cost includes both cryptographic and non-cryptographic operations. The reason for introducing this metric is to evaluate the effect of non-cryptographic operations on the protocol execution.
- **Authentication delay** is defined as the time needed for a client device to authenticate itself during an authentication instance.

The following sections describe how each experiment is conducted, how the results are collected and experimental results. We use the 1-machine set-up for Experiment-1/-2, and the 2-machine set-up for Experiment-3.

### B. EXPERIMENT-1

This experiment is to evaluate the PCC costs of our protocols.

#### 1) Experiment Setting and the Number of Iterations

The settings of the experiment are as follows:

- **Set-up:** 1-machine set-up.
- **Machine:** M1.
- **Performance metric:** PCC cost.

To ensure statistical significance of the experimental results, the number of iterations (n) over which the executions time is measured should be determined. This is done by experimenting and measuring the average execution times of three cryptographic algorithms (AES encryption, SHA256, and RSA encryption). The results are shown in Figure 15. As shown in the figure, when n is sufficiently large, e.g., $n > 5.5K$, the results show very little fluctuations, meaning they are hardly affected by system dynamics. The confidence level of the results is measured using the standard error of the mean (SEM). SEM represents the error of a sample mean. It is the standard deviation ($\sigma$) of the sample divided by the square root of n (i.e., SEM $=\sigma/\sqrt{(n)}$) [54]. The smaller the SEM value, the more representative the sample. The Experiment-1 results presented in this section are collected by using the n value of $7K$ and the corresponding SEM value is 0.007.

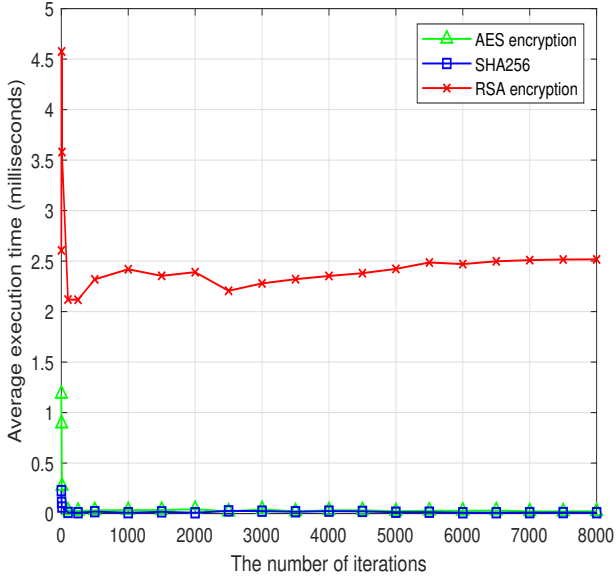

FIGURE 15: Experiment-1: Number of iterations

2) Experiment Results

Figure 16 shows the average execution times of the cryptographic algorithms. The average times are, respectively, 0.009 ms for a hash operation ($T_H$), 0.030 ms for an HMAC operation ($T_{HMAC}$), and 0.018 ms for a symmetric encryption or decryption operation ($T_{SE}$).

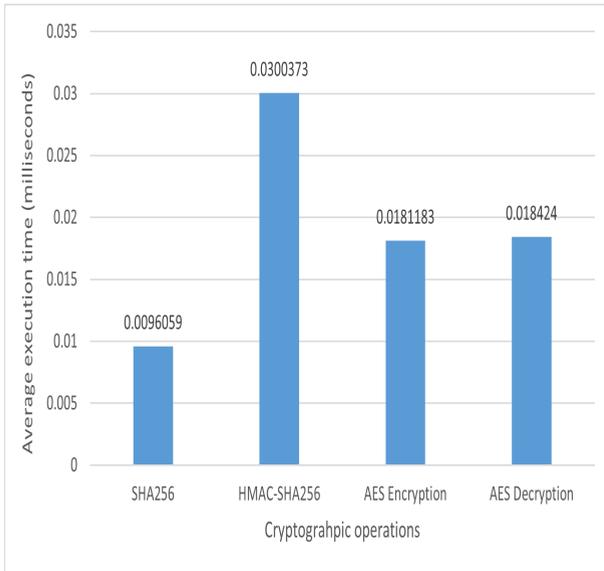

FIGURE 16: Computation costs of the cryptographic algorithms

In the theoretical evaluation (discussed in the performance evaluation of each protocol), the computation costs of the M2I protocols are evaluated in terms of the number of cryptographic operations performed and the type of cryptographic algorithms used to perform them. Here, we measure these costs in terms of the time taken to execute all the cryptographic operations during the execution of a protocol. The computation costs of the protocols are presented in Table 14. The costs are dependent on the number of devices involved in each protocol execution.

Figure 17 shows the PCC costs for authentication (also referred to as initial authentication) and re-authentication using the P2P and O2M protocol. From the figure, it can be seen that the costs for authentication in both protocols increase as the number of target devices increases. However, the rate of increase in the O2M protocol is lower than that of the P2P protocol. For example, as the number of target devices increases from 5 to 400, the PCC cost of the P2P protocol increases from 1 ms to 90 ms whereas the corresponding cost for the O2M protocol increases from 0.8 ms to 54 ms, which is 0.6 times lower. The O2M protocol is cheaper than the P2P protocol; it reduces the cost by $32\% \sim 40\%$, in comparison with the P2P protocol. This is because, with the O2M protocol, each client device uses the same token (i.e., O2M-Ticket) to access all the target devices, whereas, with the P2P protocol, each client device uses a separate token (i.e., P2P-Ticket) to connect to a different target device, so when accessing multiple target devices, multiple tokens are required. The figure also shows that the P2P and O2M protocol introduce the same level of PCC cost during re-authentication. This is because, as discussed earlier, the cryptographic operations used for re-authentication in the case of two protocols are identical.

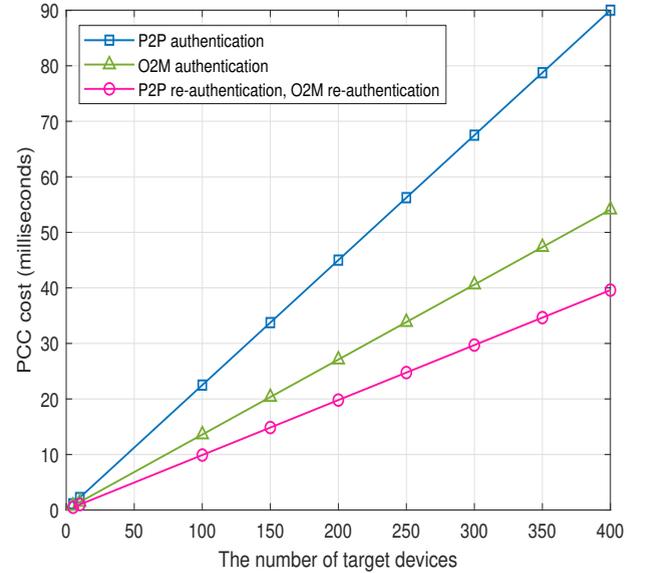

FIGURE 17: PCC costs of the P2P and O2M protocol

The average execution times of the cryptographic algorithms used in Kerberos, on a 392-bit long message (i.e., the maximum message component that uses these algorithms), are as follows. 0.055 ms for a symmetric encryption operation ($T_{KSE}$), and 0.080 ms for a symmetric decryption operation ($T_{KSD}$).

TABLE 14: Computation costs of the M2I protocols

| Protocol | Cryptographic operations | PCC cost $(ms)$ |
|---|---|---|
| P2P | $NT(12T_{SE} + T_H)$ | $0.225 \times NT$ |
| P2P re-authentication | $NT(5T_{SE} + T_H)$ | $0.099 \times NT$ |
| O2M | $5T_{SE} + NT(7T_{SE} + T_H)$ | $0.135 \times NT + 0.09$ |
| O2M re-authentication | $NT(5T_{SE} + T_H)$ | $0.099 \times NT$ |

### C. EXPERIMENT-2

This experiment is to evaluate the PTC costs of our protocols.

1) Experiment Setting and the Number of Iterations

The setting of this experiment is identical to Experiment-1, with the exception of the performance metric used to evaluate the experiment.

- **Performance metric:** PTC costs

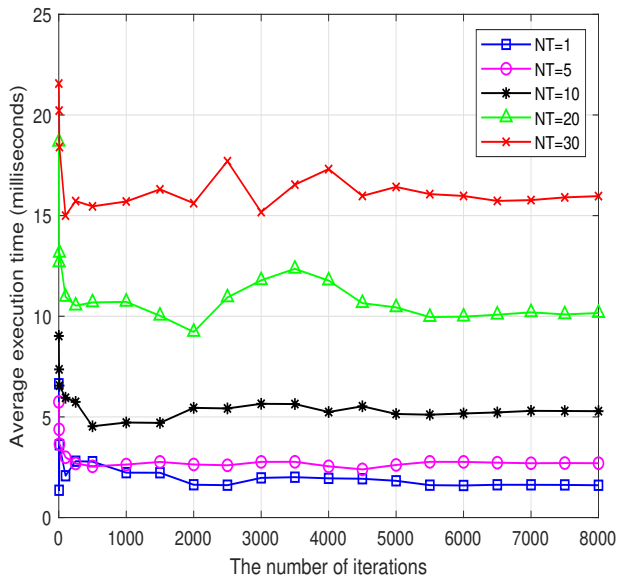

FIGURE 18: Experiment-2: Number of iterations

To identify the number of iterations (n) over which the overall execution time is measured, we ran the P2P protocol using the same method discussed in Experiment-1. Any of the M2I protocols could be used to identify n. This is because the n value is chosen based on the level of fluctuations caused by system dynamics and hence it is not dependent on a specific protocol. Figure 18 shows that when n is larger than $6.5K$, very little fluctuations occur. The Experiment-2 results are collected by using the n value of $7K$ and the corresponding SEM value is 0.02.

2) Experiment Results

Figure 19 shows the PCC cost (discussed in Experiment-1) and the PTC cost (obtained from this experiment) for authentication using the P2P protocol. From the figure, it can be seen that both costs increase as the number of target devices increases. However, the rate of increase in the PCC cost is lower than that of the PTC cost. For example, as the number of target devices increases from 5 to 100, the PCC cost of the P2P protocol increases from 1 ms to 23 ms whereas the corresponding PTC cost increases from 2 ms to 52 ms, which is two times higher. To measure the difference, we calculated the percentages of the PCC cost to the PTC cost. The percentages are: $42\%$ when NT = 250, or 300, $43\%$ when NT = 10, 100, 150, 200, 350, or 400, and $50\%$ when NT = 5. Therefore, The PCC cost of the P2P protocol is $42\% \sim 50\%$ of the PTC cost.

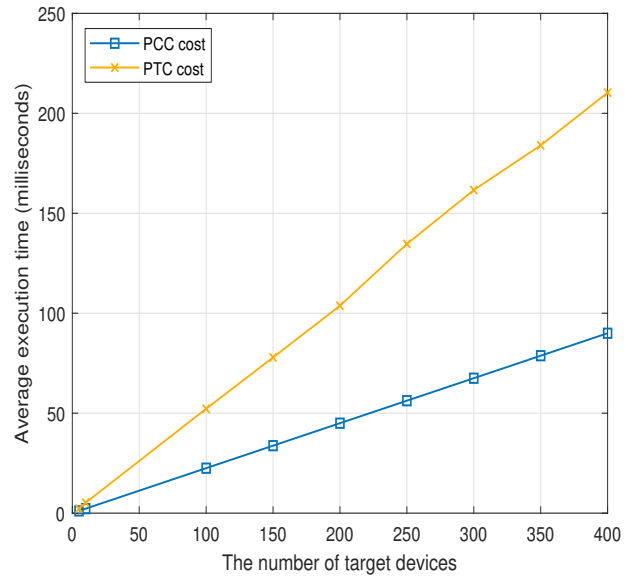

FIGURE 19: Experiment-2: P2P protocol

Figure 20 shows the PCC cost and the PTC cost for authentication using the O2M protocol. Similar to the P2P protocol, the rate of increase in the PTC cost is double the PCC cost. For example, as the number of target devices increases from 5 to 100, the PCC cost of the O2M protocol increases from 0.8 ms to 14 ms whereas the corresponding PTC cost increases from 2 ms to 26 ms, which is two times higher. The percentages of the PCC cost to the PTC cost are as follows. $47\%$ when NT = 10, $49\%$ when NT = 5, $51\%$ when NT = 200, 250, 300, or 350, and $52\%$ when NT = 100, 150, or 400. Hence, the PCC cost of the O2M protocol is $47\% \sim 52\%$ of the PTC cost.

The difference between the PCC cost and the PTC cost of the P2P and O2M protocol can be attributed to two reasons. The first is that the PTC cost measures the time required to perform all the operations, not just cryptographic but also non-cryptographic operations. An example of a non-cryptographic operation used in our protocols is the timestamp verification using the TS-Veri algorithm. On the other hand, the PCC cost only measures the time required to perform the cryptographic operations in a protocol. The second reason is that there are some additional costs introduced by the hidden operations of the P2P and O2M protocol prototypes. One such hidden operation is a byte serialisation operation. To encrypt a message component that has items of different data types, the component needs to be serialised (i.e., converted to a stream of bytes) before it can be encrypted. To read an encrypted message component that has items of different data types, the component needs to be decrypted, then, de-serialised (i.e., converted to its original data types). Also, to send a message over the network, the message needs to be serialised (if it is not in bytes format) before it can be sent, and de-serialised once it has been received.

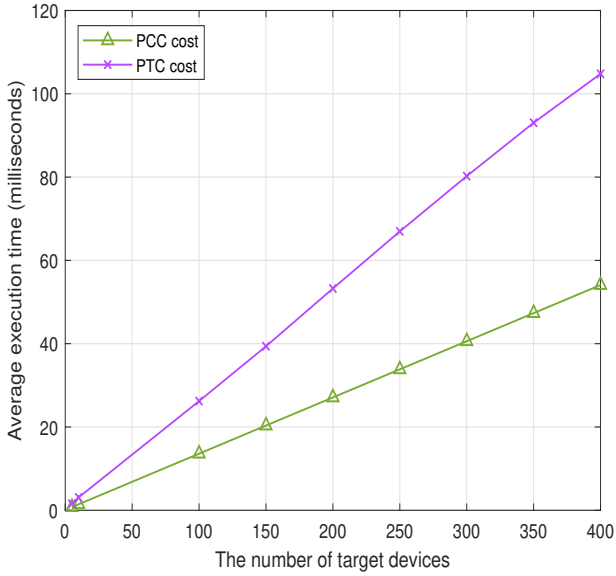

FIGURE 20: Experiment-2: O2M protocol

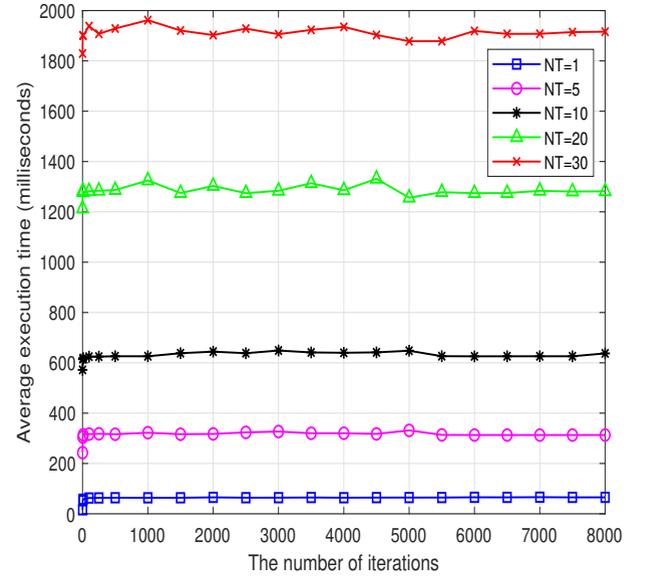

FIGURE 21: Experiment-3: Number of iterations

### D. EXPERIMENT-3

This experiment is to evaluate the authentication delays of the protocols.

#### 1) Experiment Setting and the Number of Iterations

The settings of Experiment-3 are as follows.

- **Set-up:** 2-machine set-up.

- **Machines:** M1, M2.
  M1 is used to run the client device, whereas M2 is used to run the authentication server, and target devices.

- **Performance metric:** authentication delay.

To find a suitable number of iterations for this experiment, we have measured the average authentication delays of the P2P protocol with varying iteration values. The results are shown in Figure 21. From the figure, it can be seen that the results are hardly affected by system dynamics when n is sufficiently large, e.g., $n > 6K$. The Experiment-3 results are collected by using the n value of $7K$ and the corresponding SEM value is $0.3$.

#### 2) Experiment Results

Table 15 shows the authentication delays with the P2P and O2M protocol, respectively. From the table, we can make the following observations. First, with both protocols, the authentication delays increase with an increasing number of target devices. This is expected as the number of tokens issued and verified, and the number of interactions (i.e., messages exchanged) increase as the number of target devices increases. Secondly, the authentication delays of the O2M protocol are lower than those of the P2P protocol in all the cases due to the same reasons mentioned in Experiment-1, but the rate of the reduction diminishes as the number of target devices increases. For example, when the target device number is 1, the O2M protocol reduces the delay by $14\%$ in comparison with the P2P protocol. However, when this number increases to 100, the difference between the two delays is $0.8\%$. This is because initialisation factors have less effect on the authentication delay as the number of target devices increases. Finally, the O2M protocol improves the performance by $0.2\% \sim 14\%$ in comparison with the P2P protocol. For instance, when the number of target devices (NT) is less than 11 devices, the authentication delay decreases by $6\% \sim 14\%$, and when NT is more than 99 devices, the delay decreases by $0.2\% \sim 0.8\%$.

TABLE 15: Authentication delays of the P2P and O2M protocol

| Number of target devices | Authentication delays (*seconds*) | | DD=$Delay_{P2P} - Delay_{O2M}$ ($DD/Delay_{P2P}\%$) |
|---|---|---|---|
| | P2P Protocol ($Delay_{P2P}$) | O2M Protocol ($Delay_{O2M}$) | |
| 1 | 0.07 | 0.06 | 0.01 (14%) |
| 5 | 0.36 | 0.32 | 0.04 (11%) |
| 10 | 0.54 | 0.51 | 0.03 (6%) |
| 100 | 6.46 | 6.41 | 0.05 (0.8%) |
| 150 | 9.57 | 9.55 | 0.02 (0.2%) |
| 200 | 12.71 | 12.69 | 0.02 (0.2%) |
| 250 | 15.97 | 15.90 | 0.07 (0.4%) |
| 300 | 19.18 | 19.12 | 0.06 (0.3%) |
| 350 | 22.31 | 22.25 | 0.06 (0.3%) |
| 400 | 25.42 | 25.24 | 0.18 (0.7%) |

## IX. M2I PROTOCOLS vs KERBEROS

To evaluate the efficiency of the M2I protocols, this section compares the protocols to Kerberos version 5 in terms of communication and computational costs.

### A. KERBEROS

Kerberos is a symmetric-key based authentication protocol that uses tickets to provide client/server authentication. Kerberos is a well-known and widely used authentication solution [55] that could be used to implement device multi-factor authentication. Hence, it is a suitable benchmark solution. In this section, we discuss Kerberos messages, communication and computation costs.

#### 1) Kerberos messages

Kerberos protocol consists of six messages as shown in Figure 22. The first two messages (i.e., $Msg1$ and $Msg2$) are used to authenticate a client device ($C$) to the authentication server ($AS$) to obtain a ticket-granting ticket ($K - TGT_{AS}$). The $K - TGT_{AS}$ is then used in Msg3 to authenticate $C$ to the ticket granting server ($TGS$) to get a service-granting ticket ($K - SGT_{TGS}$). Once obtained, the $K - SGT_{TGS}$ is used in $Msg5$ to authenticate $C$ to a target device ($D$). To implement two-factor authentication using Kerberos, the client device may need to obtain two service-granting tickets before sending its access request to the target device. The components of Kerberos messages are presented in Figure 23 [43].

#### 2) Performance Evaluation
##### a: Additional Assumptions

In addition to the assumptions presented in Section V-G, the following assumptions are used in the evaluation.

- Times attribute is 96-bit long. This is because it consists of three time objects (start, end, and renewal time) [43].
- Flags and options are each 32-bit long [44].
- A realm is 8-bit long [44].
- The network address is 8-bit long. This is because Class A IPv4 Internet protocol is used in this evaluation [56].

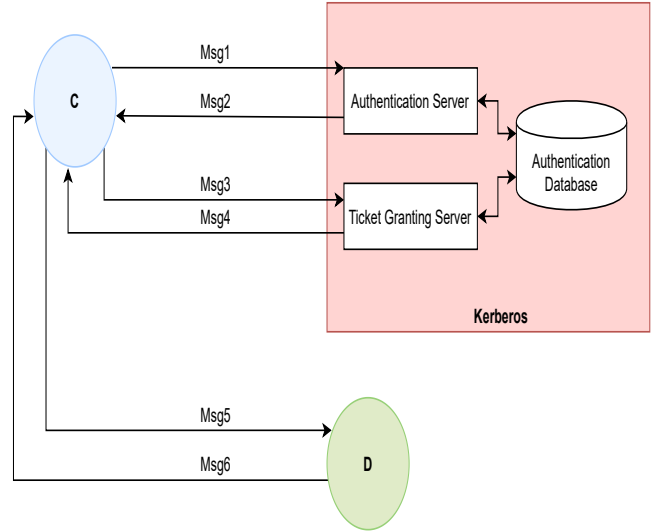

FIGURE 22: Overview of Kerberos

- The symmetric-key cipher used is the AES-128-CTS algorithm (i.e., AES-128 in Cipher Block Chaining (CBC) mode with Ciphertext Stealing, also known as CBC-CS3 mode)[57].

##### b: Communication Cost

Table 16 shows the communication costs of Kerberos tickets and authenticators. From the table, it can be seen that the costs incurred to construct a Kerberos ticket (i.e., $K - TGT_{AS}$ or $K - SGT_{TGS}$) and authenticator (i.e., $Authenticator1_{Ci}$ or $Authenticator2_{Ci}$) are 384 bits and 128 bits, respectively.

The total communication cost of one execution of Kerberos is ($2408\times$ the number of target devices ($NT$)+1264) bits for each of the client devices as presented in Table 17. For instance, if a client device authenticates itself to 3 target devices using Kerberos, the communication cost incurred would be $((2408 \times 3) + 1264)$ bits = 1061 bytes. It is worth-noting that optional message items (e.g., $Subkey$, and $Seq$) are not considered in the evaluation.

| Entities | Protocol messages | Items |
|---|---|---|
| $C_i \to$ AS | Msg1:= | $< Options\|\|ID_{C_i}\|\|Realm_{C_i}\|\|ID_{TGS}\|\|Times\|\|EnNonce1 >$ |
| AS $\to C_i$ | Msg2:= | $< Realm_{C_i}\|\|ID_{C_i}\|\|K-TGT_{AS}\|\|EK_{C_i}[K_{C_i,TGS}\|\|Times\|\|EnNonce1\|\|Realm_{TGS}\|\|ID_{TGS}] >$<br>$K-TGT_{AS} = EK_{TGS}[K-Flags\|\|K_{C_i,TGS}\|\|Realm_{C_i}\|\|ID_{C_i}\|\|AD_{C_i}\|\|Times]$ |
| $C_i \to$ TGS | Msg3:= | $< Options\|\|ID_{D_i}\|\|Times\|\|EnNonce2\|\|K-TGT_{AS}\|\|Authenticator1_{C_i} >$<br>$Authenticator1_{C_i} = EK_{C_i,TGS}[ID_{C_i}\|\|Realm_{C_i}\|\|Ts1]$ |
| TGS $\to C_i$ | Msg4:= | $< Realm_{C_i}\|\|ID_{C_i}\|\|K-SGT_{TGS}\|\|EK_{C_i,TGS}[K_{C_i,D_i}\|\|Times\|\|EnNonce2\|\|Realm_{D_i}\|\|ID_{D_i}] >$<br>$K-SGT_{TGS} = EK_{D_i}[K-Flags\|\|K_{C_i,D_i}\|\|Realm_{C_i}\|\|ID_{C_i}\|\|AD_{C_i}\|\|Times]$ |
| $C_i \to D_i$ | Msg5:= | $< Options\|\|K-SGT_{TGS}\|\|Authenticator2_{C_i} >$<br>$Authenticator2_{C_i} = EK_{C_iD_i}[ID_{C_i}\|\|Relam_{C_i}\|\|Ts2\|\|Subkey\|\|Seq]$ |
| $D_i \to C_i$ | Msg6:= | $< EK_{C_i,D_i}[Ts2\|\|Subkey\|\|Seq] >$ |

FIGURE 23: Kerberos messages

TABLE 16: Communication costs of Kerberos tickets and authenticators

| Kerberos items | Components | Total length ($bits$) |
|---|---|---|
| $K-TGT_{AS}$ | $EK_{TGS}[K-Flags\|\|K_{Ci,TGS}\|\|Realm_{Ci}\|\|ID_{Ci}\|\|AD_{Ci}\|\|Times])$ | 384 |
| $K-SGT_{TGS}$ | $EK_{Di}[K-Flags\|\|K_{Ci,Di}\|\|Realm_{Ci}\|\|ID_{Ci}\|\|AD_{Ci}\|\|Times]$ | 384 |
| $Authenticator1_{Ci}$ | $EK_{Ci,TGS}[ID_{Ci}\|\|Realm_{Ci}\|\|Ts1]$ | 128 |
| $Authenticator2_{Ci}$ | $EK_{CiDi}[ID_{Ci}\|\|Relam_{Ci}\|\|Ts2]$ | 128 |

c: Computation Cost

The total computation cost of one execution of Kerberos is $(2T_{KSE} + T_{KSD} + NT(5T_{KSE} + 6T_{KSD}))$ $ms$ as shown in Table 18. The time needed to perform these operations (as discussed in Experiment-1) is 0.055 $ms$ and 0.080 $ms$ for $T_{KSE}$ and $T_{KSD}$, respectively. Therefore, the PCC cost of the protocol is $(0.755 \times NT + 0.19)$ $ms$. For example, if a client device authenticates itself to three target devices using Kerberos, the PCC cost incurred would be $(0.755 \times 3 + 0.19) = 2.5$ $ms$.

TABLE 17: Communication cost of Kerberos

| Entities | Protocol messages | Items | Total length ($bits$) |
|---|---|---|---|
| Client device | Msg1 | $Options\|\|ID_{Ci}\|\|Realm_{Ci}\|\|ID_{TGS}\|\|$ $Times\|\|EnNonce1$ | 328 |
| AS | Msg2 | $Realm_{Ci}\|\|ID_{Ci}\|\|K-TGT_{AS}\|\|$ $EK_{Ci}[K_{Ci,TGS}\|\|Times\|\|EnNonce1\|\|$ $Realm_{TGS}\|\|ID_{TGS}]$ | 936 |
| Client device | Msg3 | $Options\|\|ID_{Di}\|\|Times\|\|EnNonce2\|\|$ $K-TGT_{AS}\|\|Authenticator1_{Ci}$ | 800 |
| TGS | Msg4 | $Realm_{Ci}\|\|ID_{Ci}\|\|K-SGT_{TGS}\|\|$ $EK_{Ci,TGS}[K_{Ci,Di}\|\|Times\|\|EnNonce2\|\|$ $Realm_{Di}\|\|ID_{Di}]$ | 936 |
| Client device | Msg5 | $Options\|\|K-SGT_{TGS}\|\|Authenticator2_{Ci}$ | 544 |
| Target device | Msg6 | $EK_{Ci,Di}[Ts2]$ | 128 |
| The total length per protocol execution | | | $2408 \times NT + 1264$ |

TABLE 18: Computation cost of Kerberos

| Kerberos | Entities | | | | Total cost ($ms$) |
|---|---|---|---|---|---|
| | Client device | Kerberos servers | | Target device | |
| | | AS | TGS | | |
| Cryptographic operations | $T_{KSD} + 2NT$ $(T_{KSE} + T_{KSD})$ | $2T_{KSE}$ | $2NT(T_{KSE} + T_{KSD})$ | $NT(T_{KSE} + 2T_{KSD})$ | $2T_{KSE} + T_{KSD} +$ $NT(5T_{KSE} + 6T_{KSD})$ |
| PCC cost ($ms$) | $0.27 \times NT + 0.080$ | 0.11 | $0.27 \times NT$ | $0.215 \times NT$ | $0.755 \times NT + 0.19$ |

### B. PERFORMANCE EVALUATION OF THE M2I PROTOCOLS vs KERBEROS

This section compares the communication and computational cost of the M2I protocols with that of Kerberos.

#### 1) Communication Costs

The total communication costs of the P2P, O2M, and Kerberos protocol, as discussed in Section VI-D2c, Section VI-D3c, and Section IX-A2, are shown in Table 19. The costs of two-factor authentication are $(4866 \times NT)$ bits, $(2 \times ((1281 \times NT) + 128[(32 \times NT) + 192] + 896))$ bits, and $((4816 \times NT) + 1264)$ bits in the P2P, O2M, and Kerberos protocol, respectively.

Figure 24 shows the communication costs of one-factor authentication using the P2P, O2M, and Kerberos protocol, respectively. From the figure, we can make the following observations. First, the costs of the protocols are similar when the number of target devices is less than 10. However, when the number of target devices goes beyond 10, the communication costs of the P2P and Kerberos protocol increase steadily as the number of target devices increases, and the rate of the increase is similar for both protocols. This is because once a client device obtained a $K-TGT_{AS}$ in its first authentication instance using Kerberos, both protocols use one ticket to authenticate the client device to a target device. Secondly, the communication cost of the O2M protocol is significantly lower than that of the P2P and Kerberos protocol. The O2M protocol reduces the communication cost by $42\% \sim 45\%$ in comparison with that of Kerberos. As explained in Experiment-1, the reason for this is that the O2M protocol allows the client device to use the same token to access all the target devices.

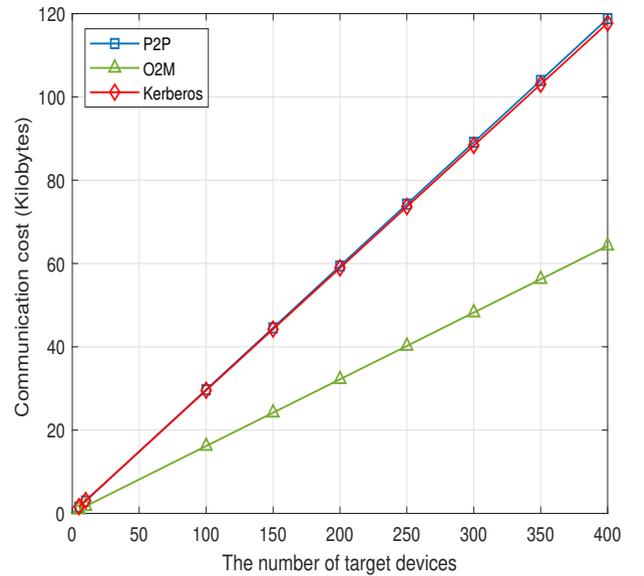

FIGURE 24: Communication costs of one-factor authentication

TABLE 19: Communication costs of the P2P and O2M protocol vs Kerberos

| Authentication type | One-factor | | | Two-factor | | |
|---|---|---|---|---|---|---|
| Protocol | P2P | O2M | Kerberos | P2P | O2M | Kerberos |
| The total length per protocol execution ($bits$) | $2433 \times NT$ | $(1281 \times NT) + 128[(32 \times NT) + 192] + 896$ | $(2408 \times NT) + 1264$ | $4866 \times NT$ | $2 \times ((1281 \times NT) + 128[(32 \times NT) + 192] + 896)$ | $(4816 \times NT) + 1264$ |

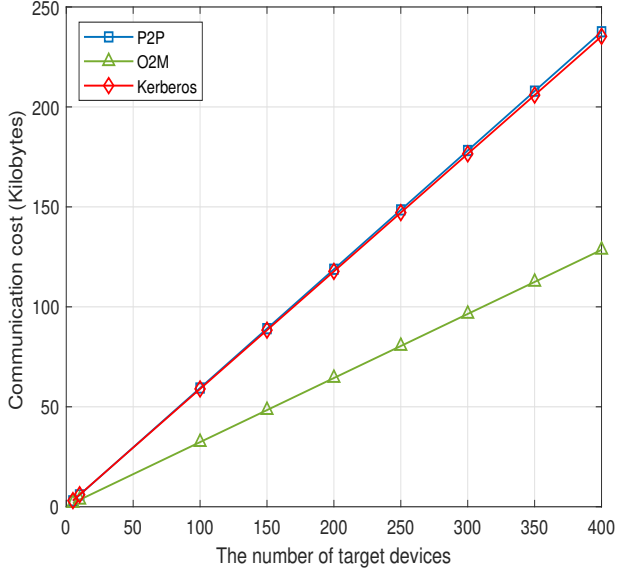

FIGURE 25: Communication costs of two-factor authentication

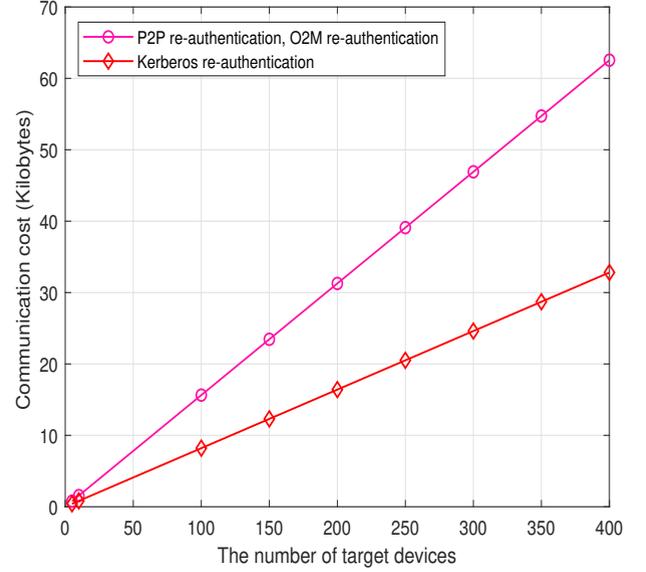

FIGURE 26: Communication costs for re-authentication

Figure 25 shows the communication costs of two-factor authentication using the P2P, O2M, and Kerberos protocol. The results shown in this figure exhibit the same patterns as those in Figure 24 – the one-factor authentication case, with the exception that the communication cost in this case doubles that of the one-factor authentication case. For example, when NT is 400, the communication costs respectively introduced by the P2P and Kerberos protocol are 238 $Kbytes$ as against 119 $Kbytes$ in the one-factor authentication case. Similarly, when NT is 400, the communication cost introduced by the O2M protocol is 128 $Kbytes$ as against 64 $Kbytes$ in the one-factor authentication case.

The communication costs for re-authentication using the protocols are presented in Table 20. The costs incurred to acquire access credentials are not considered in the evaluation. This is because it is assumed that a client device has a valid access token from its initial authentication.

TABLE 20: Communication costs for re-authentication using the P2P and O2M protocol vs Kerberos

| Protocol | P2P | O2M | Kerberos |
|---|---|---|---|
| The total length per protocol execution ($bits$) | $1281 \times NT$ | $1281 \times NT$ | $672 \times NT$ |

Figure 26 shows the communication costs for re-authentication using the protocols. From the figure, it can be seen that the cost of our protocols increases steadily, whereas the cost of Kerberos increases at a lower rate, with the increase of the number of devices. The P2P and O2M protocol, respectively, increase the communication cost by 91% in comparison with that of Kerberos. This is because the P2P and O2M protocol use hashed nonces for re-authentication, as the clocks of IoT devices may not be synchronised, whereas the Kerberos uses timestamps, and the former imposes more communication cost.

2) Computation Costs

The PCC costs of the P2P, O2M, and Kerberos protocol are shown in Table 21 and Table 22.

Figure 27 shows the PCC costs of one-factor authentication using the P2P, O2M, and Kerberos protocol. From the figure, it can be seen that the costs of our protocols increase slightly, whereas the cost of Kerberos increases steadily, at a much higher rate, with the increase of the number of devices. The P2P and O2M protocol, respectively, reduce the PCC cost by 70% ∼ 72% and 81% ∼ 82% in comparison with that of Kerberos. This is due to the higher number of tokens and the symmetric-key cipher used in Kerberos as discussed in Section IX-A.

TABLE 21: Computation costs of the P2P and O2M protocol vs Kerberos using one-factor authentication

| Authentication type | One-factor | | |
|---|---|---|---|
| Protocol | P2P | O2M | Kerberos |
| Cryptographic operations | $NT(12T_{SE} + T_H)$ | $5T_{SE} + NT(7T_{SE} + T_H)$ | $2T_{KSE} + T_{KSD} + NT(5T_{KSE} + 6T_{KSD})$ |
| PCC cost ($ms$) | $0.225 \times NT$ | $0.135 \times NT + 0.09$ | $0.755 \times NT + 0.19$ |

TABLE 22: Computation costs of the P2P and O2M protocol vs Kerberos using two-factor authentication

| Authentication type | Two-factor | | |
|---|---|---|---|
| Protocol | P2P | O2M | Kerberos |
| Cryptographic operations | $2NT(12T_{SE} + T_H)$ | $10T_{SE} + 2NT(7T_{SE} + T_H)$ | $2T_{KSE} + T_{KSD} + 2NT(5T_{KSE} + 6T_{KSD})$ |
| PCC cost ($ms$) | $0.45 \times NT$ | $0.27 \times NT + 0.18$ | $1.51 \times NT + 0.19$ |

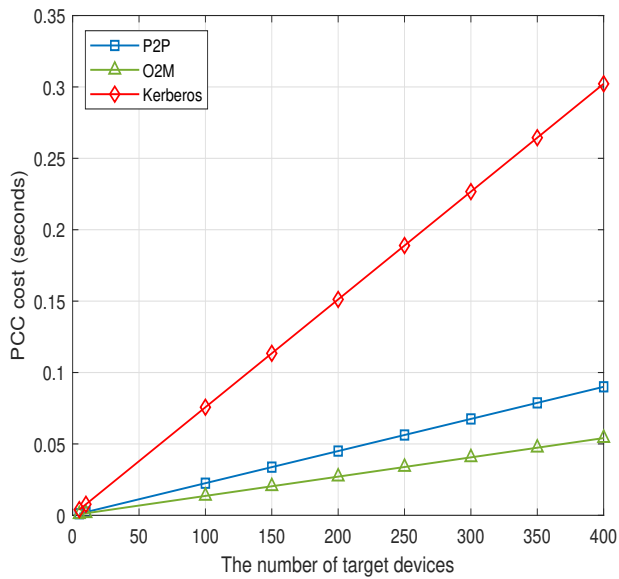

FIGURE 27: PCC costs of one-factor authentication

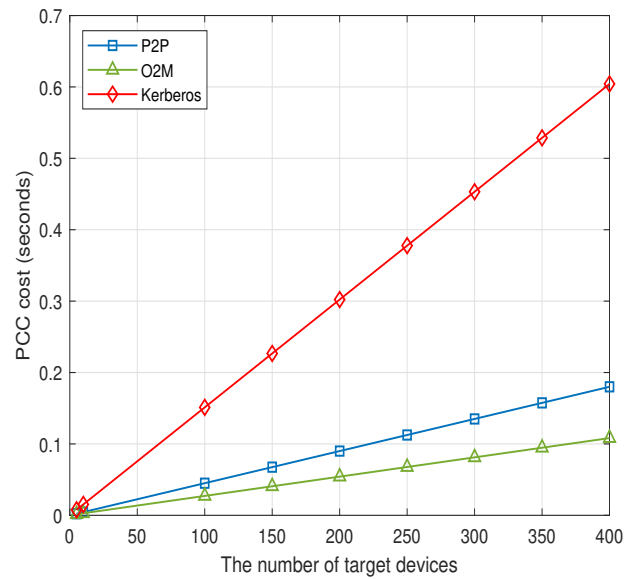

FIGURE 28: PCC costs of the two-factor authentication

Figure 28 shows the PCC costs of two-factor authentication using the P2P, O2M, and Kerberos protocol. The results shown in this figure exhibit the same patterns as those in Figure 27 – the one-factor authentication case, with the exception that the PCC cost in this case doubles that of the one-factor authentication case. For example, when NT is 400, the PCC cost introduced by Kerberos is 0.6 second as against 0.3 second in the one-factor authentication case.

The PCC costs for re-authentication using the protocols are presented in Table 23. The costs incurred to obtain access credentials are not considered in the evaluation due to the same reason mentioned in Section IX-B1.

Figure 29 shows the PCC cost for re-authentication using the P2P, O2M, and Kerberos protocol. From the figure, it can be seen that the cost of our protocols increases slightly, whereas the cost of Kerberos increases steadily, at a much higher rate, with the increase of the number of devices. The P2P and O2M protocol, respectively, reduce the PCC cost by 72% in comparison with that of Kerberos. This is due to the different symmetric-key ciphers used in the protocols. In the P2P and O2M protocol, the cipher used is the AES-128-CBC algorithm, whereas the cipher used in Kerberos is the AES-128-CTS algorithm, and the latter is much more computationally expensive as discussed in Experiment-1.

## X. CONCLUSION

Authentication solutions producing a higher assurance level provide a higher level of protection, but they often impose a higher communication and computational overhead, in comparison with authentication solutions producing a lower assurance level. Hence, there is a need to optimise the trade-off between the level of protection and overhead costs. In this paper, we have critically analysed the level of assurance required and derived during authentication, and proposed a number of methods to quantify them. The M2I framework has then been proposed to facilitate multi-LoA and interaction based authentication for IoT to reduce the costs incurred and enhance the security level of IoT applications. The M2I protocols have been evaluated in terms of security and performance. The security evaluation shows that the protocols satisfy the security requirements and are resilient to known attacks. The performance evaluation shows that using the O2M interaction mode in authentication can reduce the communication cost considerably. The O2M protocol cuts the communication cost by $42\% \sim 45\%$ compared with that of the Kerberos protocol. The evaluation also shows that the P2P and O2M protocol cut the computational cost by $70\% \sim 72\%$ and $81\% \sim 82\%$ in comparison with that of Kerberos, respectively. Hence, adopting the LoA linked and interaction-based key sharing for authentication can provide a more effective and efficient protection for IoT applications. As part of our future work, we plan to extend the M2I framework to address the remaining modes of entity interactions, such as multiDevice-to-device interactions.

TABLE 23: Computation costs for re-authentication using the P2P and O2M protocol vs Kerberos

| Protocol | P2P | O2M | Kerberos |
|---|---|---|---|
| Cryptographic operations | $NT(5T_{SE} + T_H)$ | $NT(5T_{SE} + T_H)$ | $NT(2T_{KSE} + 3T_{KSD})$ |
| PCC cost ($ms$) | $0.099 \times NT$ | $0.099 \times NT$ | $0.35 \times NT$ |

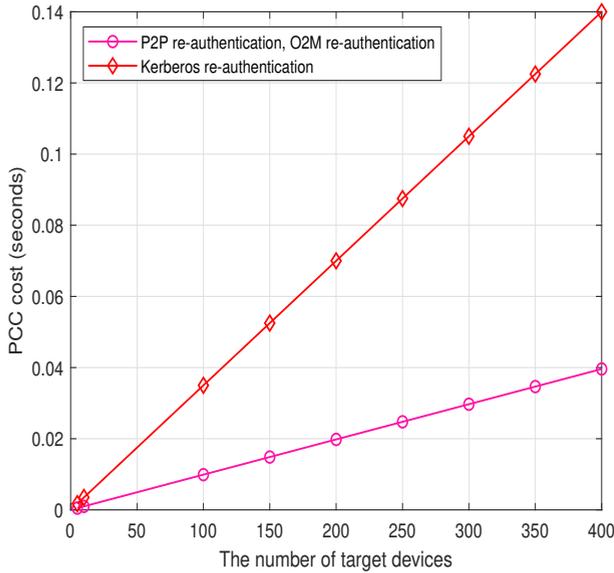

FIGURE 29: PCC costs for re-authentication

**Salem AlJanah** received the B.Sc. degree (Hons.) in information systems from Al-Imam Muhammad Ibn Saud Islamic University, Saudi Arabia, and the M.Sc. degree (Dist.) in information systems and technology from the University of Michigan, USA. He is currently pursuing the Ph.D. degree in Internet of Things (IoT) Security with The University of Manchester, U.K.

He is a CISSP and Security+ certified security professional, and a Fellow of the Higher Education Academy in the UK. His research interests include the IoT, applied cryptography, network and web security.

**Ning Zhang** received the B.Sc. degree (Hons.) from Dalian Maritime University, China, and the Ph.D. degree from the University of Kent, U.K., both in electronics engineering.

Since 2000, she has been with the Department of Computer Science, The University of Manchester, U.K., where she is currently a Senior Lecturer. Her research interests include security in networked and distributed systems, applied cryptography, data privacy, trust, and digital right managements.

**Siok Wah Tay** received the B.Sc. degree in security technology from Multimedia University, Malaysia, and the M.Sc. degree in human–computer interaction from the University of Bath, U.K. She is currently pursuing the Ph.D. degree in computer science with The University of Manchester, U.K..

Her research interests include security, the IoT, and human–computer interaction.

# APPENDIX.
## A. ALGORITHMS
The algorithms used in our protocols are as follows.

---

**Algorithm 1:** The TS-Veri algorithm
---
1: **algorithm** TS-Veri(Ts)
2: **read** $T_{now}$
3: **if** $(|Ts - T_{now}| <= \triangle T)$ **then**
4:    **return** True
5: **else**
6:    **return** False
7: **end if**
8: **end function**

---

**Algorithm 2:** The ID-Veri algorithm
---
1: **algorithm** ID-Veri($ID_S$, $ID_C$)
2: **if** $(ID_S = ID_C)$ **then**
3:    **return** True
4: **else**
5:    **return** False
6: **end if**
7: **end function**

---

**Algorithm 3:** The EN-Veri algorithm
---
1: **algorithm** EN-Veri(EnNonce1, EnNonce2)
2: **if** $(EnNonce1 = EnNonce2)$ **then**
3:    **return** True
4: **else**
5:    **return** False
6: **end if**
7: **end function**

---

**Algorithm 4:** The HC-Gen algorithm
---
1: **algorithm** HC-Gen(EnNonce, n)
2: **array** HC[n] $\leftarrow \emptyset$
3: HC[0] $\leftarrow$ Hash(EnNonce)
4: **for** i$\leftarrow$1 **to** n-1 **do**
5:    HC[i]$\leftarrow$ Hash(HC[i-1])
6: **end for**
7: **return** HC
8: **end function**

---

**Algorithm 5:** The HC-Veri algorithm
---
1: **algorithm** HC-Veri($h_0$, $h_1$)
2: **if** $(Hash(h_0) = h_1)$ **then**
3:    **return** True
4: **else**
5:    **return** False
6: **end if**
7: **end function**

---

**Algorithm 6:** The TI-Veri algorithm
---
1: **algorithm** TI-Veri($ID_S$, Ticket)
2: **if** (ID-Veri($ID_S$, Ticket.$ID_C$)) **then**
3:    **read** $T_{now}$
4:    **if** (Ticket.Start-time $<=$ $T_{now}$) **and** (Ticket.End-time $>$ $T_{now}$) **and** (Ticket.LoA $>=$ RLoA) **then**
5:      **return** True
6:    **else**
7:      **return** False
8:    **end if**
9: **else**
10:    **return** False
11: **end if**
12: **end function**

---

**Algorithm 7:** The TAuthenticator-Gen algorithm
---
1: **algorithm** TAuthenticator-Gen ($ID_C$, Ticket, EnNonce, SK)
2: **if** (Ticket.Flag[3] = Reusable) **then**
3:    **input** the number of times (n) the client intends to use the ticket
4:    $h_n \leftarrow$ Hash(EnNonce)
5:    **for** i$\leftarrow$1 **to** n-1 **do**
6:      $h_n \leftarrow$ Hash($h_n$)
7:    **end for**
8:    Authenticator $\leftarrow ID_C||h_n$
9: **else**
10:    Authenticator $\leftarrow ID_C||EnNonce$
11: **end if**
12: **return encrypt**$_{SK}$(Authenticator)
13: **end function**

**Algorithm 8:** The TAuthenticator-Veri algorithm

1: **algorithm** TAuthenticator-Veri(Ticket, Authenticator, Authentication-Type)
2: **if** (ID-Veri($Ticket.ID_C$, $Authenticator.ID_C$)) **then**
3:   **if** (Authentication-Type = initial authentication) **then**
4:     **return** True
5:   **else**
6:     **read** $h_{n-1}$ from previous authentication instance
7:     $h_{new} \leftarrow$ Hash($h_{n-1}$)
8:     **if** (Authenticator.$h_n$ = $h_{new}$) **then**
9:       **return** True
10:     **else**
11:       **return** False
12:     **end if**
13:   **end if**
14: **else**
15:   **return** False
16: **end if**
17: **end function**

## B. FORMAL VERIFICATION CODE OF THE M2I PROTOCOLS

### 1) The P2P Protocol

```
role role_Client(C:agent,D:agent,AS:agent,H:
    hash_func,Kcas:
symmetric_key,SND,RCV:channel(dy))
played_by C
def=
  local
    State:nat,
        EnNonceC1,EnNonceC2,EnNonceD1,TS1,
    Ticket_Fields:text,
        Hn:message,
        Ksk,Kdas:symmetric_key
  init
    State := 0
  transition
    1. State=0 /\ RCV(start) =|> State':=1 /\
    EnNonceC1':=new() /\ TS1':=new()/\ SND({C.D.
    EnNonceC1'.TS1'}_Kcas) /\ secret(EnNonceC1',
    sec_1,{C,D,AS})
    2. State=1 /\ RCV({Ksk'.EnNonceC1.{
    Ticket_Fields'.Ksk'.
    EnNonceC1}_Kdas'}_Kcas) =|> State':=2 /\
    EnNonceC2':=new()/\ Hn':=H(EnNonceC2')
    /\ SND({Ticket_Fields'.Ksk'.EnNonceC1}_Kdas'.{
    C.Hn'}_Ksk)
    /\ request(C,AS,c_auth_as,EnNonceC1)
    /\ witness(C,D,d_auth_c,Hn')
    3. State=2 /\ RCV({EnNonceC1.EnNonceD1'}_Ksk')
     =|> State':=3
    /\ SND({EnNonceD1'}_Ksk')
    /\ request(C,D,c_auth_d,EnNonceC1)
    /\ witness(C,D,d_auth2_c,EnNonceD1')
end role

role role_AuthenticationServer(AS:agent,C:agent,D:
    agent,Kcas,Kdas:
symmetric_key,SND,RCV:channel(dy))
played_by AS
def=
  local
    State:nat,
    EnNonceC1,S,TS1,Ticket_Fields:text,
    Ksk:symmetric_key
  init
    State := 0
  transition
    1. State=0 /\ RCV({C.D. EnNonceC1'.TS1'}_Kcas)
     =|>State':=1
    /\ Ksk':=new() /\ Ticket_Fields':=new()/\
    SND({Ksk'.EnNonceC1.{Ticket_Fields'.Ksk'.
    EnNonceC1'}_Kdas}_Kcas)
    /\ secret(Ksk',sec_2,{C,D,AS})
    /\ secret(Ticket_Fields,sec_3,{D,AS})
    /\ witness(AS,C,c_auth_as,EnNonceC1)
end role

role role_TargetDevice(D:agent,C:agent,AS:agent,H:
    hash_func,Kdas:
symmetric_key,SND,RCV:channel(dy))
played_by D
def=
```

```
  local
    State:nat,
    EnNonceC1,EnNonceD1,Ticket_Fields:text,
    Hn:message,
    Ksk:symmetric_key
  init
    State := 0
  transition
    1. State=0 /\ RCV({Ticket_Fields'.Ksk'.
    EnNonceC1'}_Kdas.
    {C.Hn'}_Ksk') =|>State':=1 /\EnNonceD1':=new()
    /\ SND({EnNonceC1'.EnNonceD1'}_Ksk')
    /\ secret(EnNonceD1',sec_4,{C,D})
    /\ request(D,C,d_auth_c,Hn')
    /\ witness(D,C,c_auth_d,EnNonceC1')
    2. State=1 /\ RCV({EnNonceD1}_Ksk) =|> State
    ':=2
    /\ request(D,C,d_auth2_c,EnNonceD1)
end role

role session(C:agent,D:agent,AS:agent,H:hash_func,
    Kcas,Kdas:
symmetric_key)
def=
  local
    SND3,RCV3,SND2,RCV2,SND1,RCV1:channel(dy)
  composition
       role_Client(C,D,AS,H,Kcas,SND1,RCV1)
    /\ role_TargetDevice(D,C,AS,H,Kdas,SND2,RCV2)
    /\ role_AuthenticationServer(AS,C,D,Kcas,Kdas,
    SND3,RCV3)
end role

role environment()
def=
  const
    kc1as,kd1as,kc2as,kd2as,kias:symmetric_key,
    c1,c2,d1,d2,as:agent,
    h:hash_func,
    sec_1,sec_2,sec_3,sec_4,c_auth_as,d_auth_c,
    d_auth2_c,
    c_auth_d:protocol_id
  intruder_knowledge = {c1,c2,d1,d2,kias}
  composition
       session(c1,d1,as,h,kc1as,kd1as)
    /\ session(c1,d2,as,h,kc1as,kd2as)
    /\ session(c2,d2,as,h,kc2as,kd1as)
    /\ session(c2,d2,as,h,kc2as,kd2as)
end role

goal
  secrecy_of sec_1
  secrecy_of sec_2
  secrecy_of sec_3
  secrecy_of sec_4
  authentication_on c_auth_as
  authentication_on d_auth_c
  authentication_on c_auth_d
  authentication_on d_auth2_c
end goal

environment()
```

2) The O2M Protocol

```
role role_Client(C:agent,D:agent,AS:agent,H:
    hash_func,Kcas:
symmetric_key,SND,RCV:channel(dy))
played_by C
def=
  local
    State:nat,
    EnNonceC1,EnNonceC2,EnNonceD1,TS1,
    Ticket_Fields:text,
    Hn:message,
    Ksk,Kgas:symmetric_key
  init
    State := 0
  transition
    1. State=0 /\ RCV(start) =|> State':=1 /\
```

```
        EnNonceC1':=new()
        /\ TS1':=new()/\ SND({C.D. EnNonceC1'.TS1'}
        _Kcas)
        /\ secret(EnNonceC1',sec_1,{C,D,AS})
    2. State=1 /\ RCV({Ksk'.EnNonceC1.{
        Ticket_Fields'.Ksk'.
        EnNonceC1}_Kgas'}_Kcas) =|> State':=2 /\
        EnNonceC2':=new()
        /\ Hn':=H(EnNonceC2')
        /\ SND({Ticket_Fields'.Ksk'.EnNonceC1}_Kgas'.{
        C.Hn'}_Ksk)
        /\ request(C,AS,c_auth_as,EnNonceC1)
        /\ witness(C,D,d_auth_c,Hn')
    3. State=2 /\ RCV({EnNonceC1.EnNonceD1'}_Ksk')
         =|> State':=3
        /\ SND({EnNonceD1'}_Ksk')
        /\ request(C,D,c_auth_d,EnNonceC1)
        /\ witness(C,D,d_auth2_c,EnNonceD1')
end role

role role_AuthenticationServer(AS:agent,C:agent,D:
    agent,Kcas,Kdas,
Kgas:symmetric_key,SND,RCV:channel(dy))
played_by AS
def=
  local
    State:nat,
    EnNonceC1,S,TS1,Ticket_Fields:text,
    Ksk:symmetric_key
  init
    State := 0
  transition
    1. State=0 /\ RCV({C.D. EnNonceC1'.TS1'}_Kcas)
         =|>State':=1
        /\ Ksk':=new() /\ Ticket_Fields':=new()
        /\ SND({Ksk'.EnNonceC1.{Ticket_Fields'.Ksk'.
        EnNonceC1'}_Kgas}_
        Kcas)
        /\ secret(Ksk',sec_2,{C,D,AS})
        /\ secret(Ticket_Fields,sec_3,{D,AS})
        /\ witness(AS,C,c_auth_as,EnNonceC1)
end role

role role_TargetDevice(D:agent,C:agent,AS:agent,H:
    hash_func,Kdas,
Kgas:symmetric_key,SND,RCV:channel(dy))
played_by D
def=
  local
    State:nat,
    EnNonceC1,EnNonceD1,Ticket_Fields:text,
    Hn:message,
    Ksk:symmetric_key
  init
    State := 0
  transition
    1. State=0 /\ RCV({Ticket_Fields'.Ksk'.
        EnNonceC1'}_Kgas.
        {C.Hn'}_Ksk') =|>State':=1 /\EnNonceD1':=new()
        /\ SND({EnNonceC1'.EnNonceD1'}_Ksk')
        /\ secret(EnNonceD1',sec_4,{C,D})
        /\ request(D,C,d_auth_c,Hn')
        /\ witness(D,C,c_auth_d,EnNonceC1')
    2. State=1 /\ RCV({EnNonceD1}_Ksk) =|> State
        ':=2
        /\ request(D,C,d_auth2_c,EnNonceD1)
end role

role session(C:agent,D:agent,AS:agent,H:hash_func,
    Kcas,Kdas,Kgas:
symmetric_key)
def=
  local
    SND3,RCV3,SND2,RCV2,SND1,RCV1:channel(dy)
  composition
      role_Client(C,D,AS,H,Kcas,SND1,RCV1)
     /\ role_TargetDevice(D,C,AS,H,Kdas,Kgas,SND2,
     RCV2)
```

```
    /\ role_AuthenticationServer(AS,C,D,Kcas,Kdas,
    Kgas,SND3,RCV3)
end role

role environment()
def=
  const
    kc1as,kc2as,kd1as,kd2as,kd3as,kgdas,kias:
    symmetric_key,
    c1,c2,d1,d2,d3,as:agent,
    h:hash_func,
    sec_1,sec_2,sec_3,sec_4,c_auth_as,d_auth_c,
    d_auth2_c,
    c_auth_d:protocol_id
  intruder_knowledge = {c1,c2,d1,d2,d3,kias}
  composition
        session(c1,d1,as,h,kc1as,kd1as,kgdas)
    /\ session(c1,d2,as,h,kc1as,kd2as,kgdas)
    /\ session(c1,d3,as,h,kc1as,kd3as,kgdas)
    /\ session(c2,d2,as,h,kc2as,kd2as,kgdas)
end role

goal
  secrecy_of sec_1
  secrecy_of sec_2
  secrecy_of sec_3
  secrecy_of sec_4
  authentication_on c_auth_as
  authentication_on d_auth_c
  authentication_on c_auth_d
  authentication_on d_auth2_c
end goal

environment()
```